\documentclass[12pt]{article}

\input epsf
\usepackage{amssymb,amsmath,wrapfig,subfigure,float}
\usepackage[matrix,arrow,curve]{xy}
\usepackage{cite}
\usepackage{multirow}
\usepackage{graphicx,color}
\usepackage[debug,pageanchor=false]{hyperref}
\definecolor{link}{rgb}{.8,.15,.1}
\hypersetup{colorlinks=true,linkcolor=link,citecolor=link,urlcolor=link,linktocpage}

\makeatletter
\@addtoreset{equation}{section}
\makeatother

\def\nn {{\cal N}}

\begin{document}

\begin{titlepage}

\begin{center}

\vskip .3in \noindent

{\Large \bf{Attractors, black objects, and holographic RG flows \\ in 5d maximal gauged supergravities \\\vspace{.2cm}}}

\bigskip

	Kiril Hristov, Andrea Rota\\

       \bigskip
		 Dipartimento di Fisica, Universit\`a di Milano--Bicocca, I-20126 Milano, Italy\\
       and\\
       INFN, sezione di Milano--Bicocca,
       I-20126 Milano, Italy

       \vskip .5in
           
       {\bf Abstract }
        \vskip .1in        
        \end{center}         
       
		We perform a systematic search for static solutions in different sectors of 5d $\nn=8$ supergravities with compact and non-compact gauged R-symmetry groups, finding new and listing already known backgrounds. Due to the variety of possible gauge groups and resulting scalar potentials, the maximally symmetric vacua we encounter in these theories can be Minkowski, de Sitter, or anti-de Sitter. There exist BPS and non-BPS near-horizon geometries and full solutions with all these three types of asymptotics, corresponding to black holes, branes, strings, rings, and other black objects with more exotic horizon topologies, supported by $U(1)$ and $SU(2)$ charges. The asymptotically AdS$_5$ solutions also have a clear holographic interpretation as RG flows of field theories on D3 branes, wrapped on compact 2- and 3-manifolds.
       

\noindent

\vfill
\eject

\end{titlepage}

\section{Introduction} 
\label{sec:intro}
Maximal gauged supergravity in five dimensions is certainly a very complex and challenging theory. Understanding its spectrum of solutions has already led to important insights into quantum gravity and string theory, in particular via the AdS/CFT correspondence. Our goal in this paper is to systematize and extend the known static solutions, potentially providing new interesting gravitational systems that are yet to be understood, holographically or otherwise.

Although $\nn=8$ was formulated back in 80's \cite{Gunaydin:1984qu,Pernici:1985ju,Gunaydin:1985cu}, 
very few analytic solutions have been constructed explicitely in this theory \cite{Cvetic:2009id}, since most of them are also solutions to the less supersymmetric truncations, which are more straightforward to approach \cite{Romans:1985ps,London:1995ib,Behrndt:1998ns,Chamseddine:1999xk,Klemm:2000nj,Chamseddine:2001hk,Gutowski:2004ez,Gutowski:2004yv,Chong:2005hr,Radu:2006va,Kunduri:2006ek,Kunduri:2007qy,Bellorin:2008we,Brihaye:2012nt}. Restricting to the $\nn=4$ or $\nn=2$ gauged supergravities however
does not allow to explore the full space of vacua of $\nn=8$, 
which can be extremely rich due to its large gauge group. The maximal compact gauge group of $\nn=8$ is $SO(6)$ and has a clear M-theory origin as a compactification on $S^5$ \cite{Cvetic:2000nc}. More generally there are versions of $\nn=8$ with a gauge group $SO(p,6-p)$ for any integer $p$ \cite{Gunaydin:1985cu}, as well as other smaller gauge groups that can be accomodated in maximal supergravity by the embedding tensor formalism \cite{deWit:2004nw}. In addition it is possible to explore many inequivalent sectors of the theory corresponding to the spontaneous breaking of $SO(6)$ or $SO(p,6-p)$ to any of its subgroups, leading to even larger spectrum of solutions.

Since AdS$_5$ is the maximally supersymmetric vacuum of the $SO(6)$ gauged supergravity, some of the most interesting types of solutions we can look for in this theory are asymptotically AdS$_5$ black holes 
and their possible near horizon geometries. It is also possible to have black branes, black strings and black objects with other horizon topologies, 
all of which can be charged under either abelian or nonabelian subgroups of $SO(6)$. Many of these black 
objects are already known in the literature in different context and under different names, 
since they correspond to RG flows of D3 branes wrapped on compact surfaces
 \cite{Maldacena:2000mw,Nieder:2000kc,Naka:2002jz,Cucu:2003yk,Gauntlett:2003di,Kim:2005ez,Gauntlett:2007ph,Benini:2013cda}. 

Our search for static solutions however does not stop with AdS asymptotics as we also consider the noncompact gaugings. 
Some of those theories have dS$_5$ and Minkowski as maximally symmetric vacuum, 
suggesting the existence of asymptotically flat or dS black hole like solutions. 
Indeed we find classes of black holes and (st)rings that were previously not known to exist as solutions of gauged supergravity. 

Many interesting results of our work concern attractors geometries for black objects. 
We find all possible solutions of this type and perform a careful analysis of their moduli space,
highlighting the important subset of BPS attractors, which can always be connected to infinity with a numerical solution of the BPS equations.
These solutions are particularly interesting for holographic applications since supersymmetry ensures that a number of physical quantities remain the same at weak and strong coupling. 
From a practical point of view this is often the key to the precise holographic description of gravitational systems. 

Before describing in more detail the different solutions we have found, 
a more qualitative discussion about the type of black objects and attractors is in order. 
In 5d extremal static black objects can have two main types of near-horizon geometries, 
the direct product spacetimes AdS$_3 \times \Sigma^2$ and AdS$_2 \times \Sigma^3$. $\Sigma^q$ is typically $S^q, \mathbb{R}^q, H^q$ (or their quotients). 
For asymptotically flat or de Sitter solutions, only the spherical topologies are allowed horizons, 
while asymptotically locally AdS$_5$ spacetimes admit solutions with all possible topologies for $\Sigma^q$ without exception. 

\begin{table}
\begin{center} \label{1}
\caption{Black objects in 5d}
\begin{tabular}{|c|c|c|} 
\hline
\begin{minipage}[c][.05\textheight]{.2\textwidth}
\begin{center}
Name
\end{center}
\end{minipage}
&
\begin{minipage}[c][.05\textheight]{.3\textwidth}
\begin{center}
Horizon
\end{center}
\end{minipage}
&
\begin{minipage}[c][.05\textheight]{.2\textwidth}
\begin{center}
Asymptotics
\end{center}
\end{minipage}
\\
\hline

Black hole (BH)
&
AdS$_2 \times$S$^3$

&

Mink$_5$, AdS$_5$, dS$_5$

\\

BH in dS
&
Mink$_2 \times$S$^3$\ , dS$_2 \times$S$^3$

&
 dS$_5$

\\

Black Brane (BB)

&

AdS$_2 \times \mathbb{R}^3$

&

AdS$_5$

\\

Toroidal BH

&

AdS$_2 \times$T$^3$

&

AdS$_5$

\\

Hyperbolic BH

&

AdS$_2 \times$H$^3$

&

AdS$_5$

\\

Topological BH

&

AdS$_2 \times ($H$^3 / \sim )$

&

AdS$_5$

\\
\hline

Black String (BS)

&

AdS$_3 \times$S$^2$

&

Mink$_5$, AdS$_5$, dS$_5$

\\

BS in dS

&

Mink$_3 \times$S$^2$\ , dS$_3 \times$S$^2$

&

dS$_5$

\\

Black Ring (BR)

&

BTZ$\times$S$^2$

&

Mink$_5$

\\

Black Brane (BB)

&

AdS$_3 \times \mathbb{R}^2$

&

AdS$_5$

\\

Toroidal BS

&

AdS$_3 \times$T$^2$

&

AdS$_5$

\\

Hyperbolic BS

&

AdS$_3 \times$H$^2$

&

AdS$_5$

\\

Higher genus BS

&

AdS$_3 \times ($H$^2 / \sim )$

&

AdS$_5$

\\
\hline

\end{tabular}\end{center}
\end{table}

Since the etymology of the black objects comes from their horizon topology, a large number of different names is generated. One can find a short dictionary between names and horizon topologies in table $1$\footnote{Solutions with AdS$_2 \times \mathbb{R}^3$ and AdS$_3 \times \mathbb{R}^2$ have the same horizon topology $\mathbb{R}^3$. Therefore, even if distinct, all such spacetimes deserve the name black branes.} \footnote{The distinction between black strings and black rings is a notable exception of the general rule, as it arises from the global embedding of the near-horizon geometry in the full spacetime (black rings are also special in the fact that they exist only in flat space) \cite{Emparan:2006mm}. One can then say that black strings have AdS$_3$ factors near their horizon, while for black rings the AdS is substituted by an extremal BTZ black hole. Locally there is no difference between the two and therefore equations of motion and BPS variations are blind to this distinction.}. Alternatively, 
instead of using the black object terminology, one can think of the asymptotically AdS solutions as RG flows of D3 branes wrapped on $\Sigma^q$. The asymptotic region then corresponds to the UV, while the horizon is the IR fixed point of the flow, where the field theory on the D3 brane becomes (super)conformal.

In this work we find background solutions, corresponding to all near-horizon geometries listed\footnote{Note that we do not consider possible punctures of the internal space. Allowing for extra sources on particular points might allow for more general near-horizon geometries \cite{Gaiotto:2009gz}.} in table $1$. In many cases we are also able to present the full black object solution that represents the flow between the near-horizon geometries and the asymptotic vacua. We look at several sectors of $\nn=8$ theories with scalars, $U(1)$ and $SU(2)$ gauge fields, chosen such that they contain the most general type of static solutions in 5d supergravity. As will be explained carefully in the next section, any other inequivalent choice of 5d supergravity theory will not lead to new types of static vacua. The variety of bosonic truncations we have allow us to embed in supergravity also some more exotic solutions found in Einstein-Yang-Mills-scalar theories, such as \cite{Winstanley:1998sn,Okuyama:2002mh,Radu:2004gu,
Brihaye:2006xc} and others.

We exhaust all possible near-horizon geometries in the theories of consideration, extending already known partial results and finding new solutions. Concerning the full flows, our main results can be summarized concisely ordered by their asymptotics:
\begin{itemize}
	\item AdS$_5$ - We present new nonabelian hyperbolic and topological black holes, supported by two $SU(2)$ gauge fields, based on a metric ansatz similar to the one in \cite{Cvetic:2009id,Okuyama:2002mh}. These solutions are non-BPS and are further extended to non-extremal solutions of arbitrary temperature, thus of potential interest for condensed matter applications. 
	
	\item Mink$_5$ - We give black string solutions that are supersymmetric near the horizon (in gauged supergravity). From the point of view of ungauged supergravity these solutions are non-BPS with supersymmetric asymptotics. This opens up the possibility for existence of extremal non-BPS black rings with supersymmetric horizons, in analogy to non-BPS black holes in 4 dimensions \cite{Hristov:2012nu}.  
	
	\item dS$_5$ - We find the embedding of extremal and thermal black holes with both abelian and nonabelian charges. To our best knowledge these solutions represent the first black hole geometries in de Sitter that can be fully embedded in a supergravity theory.
 
\end{itemize}

\noindent The plan of the paper is as follows. In section \ref{sec:sugra} we discuss in detail the main aspects of $\nn=8$ supergravity, concentrating on the scalar sectors and the allowed gaugings. Next, in section \ref{sec:vacua}, we define the bosonic truncations we are interested in and present the relevant lagrangians and equations of motion. The black hole oriented reader can directly read this section and skip many of the group theory considerations needed for understanding of the supergravity explained in section \ref{sec:sugra}. In section \ref{sec:exact} we solve the equations of motion for attractor geometries of the form $M_2 \times M_3$ with $M_2$ and $M_3$ maximally symmetric space(time)s or their quotients. This fully exhausts all near-horizon geometries of  table $1$. We find large classes of solutions and determine the subspaces in parameter space that lead to supersymmetry in section \ref{sec:RGflows}. In section \ref{sec:analytic} we briefly present full analytic 
black object solutions and comment on existing/possible numeric flows that generalize them. We finish with a summary of the main results in section \ref{sec:summary}.

\section{Maximal gauged supergravity in five dimensions}
\label{sec:sugra}

\subsection{The ungauged theory}
\label{subsec:ungauged}

Historically $\nn=8$ $d=5$ gauged supergravity was first conjectured to arise from the maximally supersymmetric 
compactification of the chiral ten-dimensional supergravity on the five sphere. 
The massless spectrum coming from this compactification was analyzed 
and organized into a single "massless" $\nn=8$ anti-de Sitter supermultiplet \cite{Gunaydin:1984fk} \cite{Kim:1985ez}, 
and later the goal of giving a complete lagrangian description was achieved in \cite{Gunaydin:1984qu}.
We will now briefly review what these authors did,  
starting from the maximal ungauged supergravity in five dimensions \cite{Cremmer:1980} 
and following a standard gauging procedure which had already been explored in four \cite{deWit:1982ig} 
and seven dimensions \cite{Pernici:1984xx}.

In five dimensions the maximally extended supersymmetry algebra in Minkowsky space has the form:
\begin{displaymath}
\{ \bar{Q}^a,Q^b \} = \Omega^{ab} P,
\end{displaymath} 
where the indices $a,b=1,..,8$ run over the fundamental representation of the $R$-symmetry group $Usp(8)$, 
the group of symplectic rotations that leave the symplectic metric $\Omega$ invariant. 
$\Omega$ and its inverse $\Omega_{ab}=-\Omega^{ab}$ are used to raise and lower $Usp(8)$ indices as:
\begin{displaymath}
Q_a = \Omega_{ab} Q^b, \ \ Q^a = \Omega^{ab} Q_b \ .
\end{displaymath} 
It is a well known fact that one cannot impose neither Weyl nor Mayorana conditions on a five dimensional spinor, 
however one can still choose symplectic Mayorana spinors:
\begin{displaymath}
Q^a = \gamma_5 Q^*_a .
\end{displaymath} 

All the fields of maximal supergravity in five dimensions belong to a single $Usp(8)$ gravity supermultiplet, 
consisting of one graviton $g_{\mu \nu}$, eight gravtini $\Psi_{\mu}^a$, 27 vector fields $A_{\mu}^{AB}$, 
48 gaugini $\chi^{abc}$ and 42 scalars $V_{AB}^{\  ab}$. 
Notice that the fermionic fields carry $Usp(8)$ indices only: 
the eight gravitini transform in the fundamental representation, 
while the gaugini transform as a rank three anti-symmetric tensor satisfying a tracelessness condition 
with respect to the symplectic metric:
\begin{displaymath}
\Omega_{ab}\chi^{abc} = 0 .
\end{displaymath} 
Besides $Usp(8)$, the full bosinic symmetry group of the ungauged theory 
also contains a global $E_6$ as $U$-duality group. 
The 27 vector fields transform in the fundamental representation of $E_6$, 
which we describe as a rank 2 antisymmetric tensor $A^{AB}$, with $A,B=1,...,8$, 
satisfying a tracelessness condition:
\begin{displaymath}
\Omega_{AB}A^{AB} = 0 .
\end{displaymath} 
The reason for this choice is that the fundamental representation of $E_6$ 
corresponds to a rank 2 antisymmetric traceless tensor of its subgroup $Usp(8)$.
\noindent $E_6$ fundamental indices are raised and lowered by complex cojugation:
\begin{displaymath}
A_{AB} \equiv (A^{AB})^*, 
\end{displaymath} 
and the reality condition reads:
\begin{displaymath}
A_{AB}=\Omega_{AC}\Omega_{BD} A^{CD}.
\end{displaymath} 
There are two possible ways of describing the scalar degrees of freedom, 
either as a group valued quantity or as an algebra-valued quantity.
Since the 42 scalars parametrize the coset $E_6/Usp(8)$, 
they transform with left and right multiplication under $E_6$ and $Usp(8)$ respectively. 
These transformation properties are made explicit if we choose 
to represent them as a group matrix, a $27 \times 27$ matrix $V_{AB}^{\  ab}$ called 27-bein.
Alternatively the scalars can be represented as a rank 4 antisymmetric traceless $Usp(8)$ tensor $P^{abcd}$ 
belonging to the algebra. This is achieved by decomposing the adjoint reprenentation of $E_6$ into $Usp(8)$ 
representations as:
\begin{equation} \label{78=42+36}
78 \rightarrow 36 \oplus 42 ,
\end{equation}
\noindent where $36$ corresponds to the $Usp(8)$ algebra and $42$ is its orthogonal part in the coset, 
corresponding to the non-compact generators. As always, starting from a group element $V$ 
one can construct an element of the algebra as $V^{-1} dV$. 
Schematically, we can write the decomposition \eqref{78=42+36} as $V^{-1} dV= Q+P$, 
or in indices:
 \begin{equation} \label{pungauged}
  \tilde{V}^{AB}_{\ \ \ cd}\  \partial_{\mu} V_{AB}^{\ \ ab} = 2 Q_{\mu[c}^{\ \ [a}\delta_{d]}^{\  b]} + P_{\mu\ \ cd}^{\ ab} .
 \end{equation}
We introduced the inverse $27$-bein $\tilde{V}$ which belongs to the $\bar{27}$ representation of $E_6$, 
the dual of the $27$. We do not report the full $\nn=8$ $E_6 \times Usp(8)$ invariant lagrangian and its supersymmetry variations, 
which can be found in \cite{Cremmer:1980}, \cite{Gunaydin:1985cu}.

\subsection{Maximal compact and noncompact gaugings}
\label{subsec:gauging}
Starting from the ungauged $\nn=8$ supergravity it is possible to perform many compact and noncompact gaugings corresponding to different
subgroups of the global symmetry $E_6$ \cite{deWit:2004nw}. 
There are $27$ vector fields $A_{AB}$ in the fundamental of $E_6$ which can be used to perform the gauging with the following criterion: 
in order to get a maximal gauging, we look for the largest subgroup $H \in E_6 $ such that its adjoint representation is contained in the $27$ of $E_6$.
This subgroup can be $SO(6)$ or any of its noncompact versions $SO(N,6-N)$, under which the $27$ of $E_6$ decomposes as: 
\begin{equation} \label{78=42+36}
27 \rightarrow 15 \oplus 6 \oplus 6,
\end{equation}
corresponding to the adjoint and two copies of the fundamental of $SO(N,6-N)$. We label the $15$ adjoint vectors as $A^{IJ}=-A^{JI}$,
where $I,J={1,..,6}$ runs over the fundamental of $SO(N,6-N)$. The remaining $6\oplus 6$ gauge fields can be labelled as $A^{I\alpha}$,
where $\alpha=\{1,2\}$ runs over the fundamental representation of $SL(2)$, an extra residual global symmetry in the gauged theory
\footnote{The gauging procedure breaks the full global symmetry $E_6$ of the ungauged theory to its subgroup $SO(N,6-N) \times  SL(2)$.}.
In order to construct a gauge invariant lagrangian it is necessary to dualize the fields $A^{I\alpha}$ to rank-two tensor fields $B^{I\alpha}$ 
such that:     
\begin{displaymath}
\ast dA = dB.
\end{displaymath}
The B-field describes the same number of degress of freedom, 
but it has no abelian gauge invariance and thus it can couple to the Yang-Mills fields $A^{IJ}$. 
It also satisfies the following first order equation of motion:
\begin{displaymath}
 B = \ast d B .
\end{displaymath}
Moving to the scalar sector, we can apply the same decomposition to the $27$-bein:
\begin{displaymath}
V_{AB}^{\  ab} = ( V_{IJ}^{\  ab} , V_{I\alpha}^{\  ab}). 
\end{displaymath}
This also allows to assign $Usp(8)$ indices to the Yang-Mills field strengths as: 
\begin{displaymath}
F_{\mu \nu}^{\  ab} \equiv  F_{\mu \nu}^{\  IJ}  V_{IJ}^{\  ab}. 
\end{displaymath}
The next step for constructing a $Usp(8) \times SO(N,6-N)$ invariant lagrangian is the introduction of covariant derivatives. 
Given a field $X_{a I}$ transforming in the fundamental of $Usp(8) \times SO(N,6-N)$ we have:
\begin{equation}
D_{\mu} X_{ a I} = \partial_{\mu} X_{ a I} + Q_{\mu a}^{\  b} X_{ b I} - g A_{\mu IJ}\eta^{JK} X_{ a K}, 
\end{equation}
where $Q$ and $A$ correspond the $Usp(8)$ and $SO(N, 6-N)$ connnections respectively.  
The tensor \eqref{pungauged} describing the scalar degrees of freedom as an algebra valued object is now defined through covariant derivatives as:
 \begin{equation} 
  \tilde{V}^{AB\ ab}  D_{\mu} V_{AB}^{\ \ cd} =  P_{\mu}^{\ abcd}.
 \end{equation}
We do not report the full gauge invariant lagrangian and supersymmetry transformations which can be found in \cite{Gunaydin:1985cu}. 
What is most interesting for discussing the space of vacua is that, after the gauging procedure, in order to restore invariance under supersymmetry 
it is necessary to introduce a scalar potential $P$ of order $g^2$ in the lagrangian. 
The scalar potential is essentially what characterizes the different gaugings: 
it can admit AdS or dS vacua as extrema, while in some cases there are no extrema at all.
All the possible compact and noncompact maximal gaugings with the corresponding extrema of the potentials are summarized in table $2$.

\begin{table} \begin{center} \label{2} \caption{Maximal gaugings and extrema of the potentials}
\begin{tabular}{|c|c|c|c|c|} 
\hline
\begin{minipage}[c][.08\textheight]{.2\textwidth}
\begin{center}
Gauge Group
\end{center}
\end{minipage}
&
\begin{minipage}[c][.08\textheight]{.15\textwidth}
\begin{center}
$SO(6)$
\end{center}
\end{minipage}
&
\begin{minipage}[c][.08\textheight]{.15\textwidth}
\begin{center}
$SO(5,1)$
\end{center}
\end{minipage}
&
\begin{minipage}[c][.08\textheight]{.15\textwidth}
\begin{center}
$SO(4,2)$
\end{center}
\end{minipage}
&
\begin{minipage}[c][.08\textheight]{.15\textwidth}
\begin{center}
$SO(3,3)$
\end{center}
\end{minipage}
\\
\hline

\begin{minipage}[c][.08\textheight]{.2\textwidth}
\begin{center}
Extremum
\end{center}
\end{minipage}
&
\begin{minipage}[c][.08\textheight]{.15\textwidth}
\begin{center}
AdS$_5$
\end{center}
\end{minipage}
&
\begin{minipage}[c][.08\textheight]{.15\textwidth}
\begin{center}
$\times$
\end{center}
\end{minipage}
&
\begin{minipage}[c][.08\textheight]{.15\textwidth}
\begin{center}
$\times$
\end{center}
\end{minipage}
&
\begin{minipage}[c][.08\textheight]{.15\textwidth}
\begin{center}
dS$_5$
\end{center}
\end{minipage}
\\
\hline

\end{tabular}
\end{center}
\end{table}
For a given gauging $SO(N,6-N)$, the scalar potential has a different form in each $H\subset SO(N,6-N)$ invariant sector.
Hence, also restricting to a given gauging, it is possible to explore many inequivalent sectors in the space of vacua, 
which makes it clear that we can produce a great variety of solutions.

\subsection{Scalar potentials and symmetry breaking}
\label{subsec:scalar}
As mentioned above, 
the fundamental object to describe the space of vacua of maximal gauged supergravity is the scalar potential.
Given the highly nontrivial structure of the scalar manifold, which is described by the quotient $E_6/Usp(8)$, 
the most general form of the scalar potential in each gauging is very complicated and requires a long discussion, which can be found in \cite{Gunaydin:1985cu} .
However for a given gauging we can select a subgroup $H \subset SO(N,6-N)$ and restrict with a consistent truncation 
to the sector of the scalar manifold that is $H$ invariant.
This leads to great simplifications and makes it possible to compute the scalar potential explicitely.

As first step we need to know how the $42$ scalar degrees of freedom can be organized into representations of the gauge group $SO(N,6-N)$.
We can start decomposing the adjoint representation of $E_6$, which has dimension $78$,
noting that the $36$ scalars corresponding to the compact generators of $Usp(8)$ are unphysical and can be set to zero. 
The adjoint representation of $E_6$ decomposes under its maximal subgroup $SL(6)\times SL(2)$ as:
\begin{equation}
78 \rightarrow (35,1) \oplus (1,3) \oplus (20',2),
\end{equation}
\noindent where $35$ and $20'$ correspond to the adjoint of $SL(6)$ and to a rank three antisymmetric tensor respectively, 
while $2$ and $3$ label the fundamental and the adjoint of $SL(2)$. 
The first great simplification consists on restricting our attention to the first of these three sectors, 
namely we keep turned on only the $35$ scalars that can be described as a $6\times 6$ traceless matrix.  
This is an allowed choice since it turns out to lead to a consistent truncation. 
The $35$ can be further decomposed into its symmetric and antisymmetric parts, containing $20$ and $15$ degrees of freedom respectively. 
Of course the $15$ scalars correspond to the adjoint of $SO(6)$ are thus unphysical, so we set them to zero. 
We are left with $20$ scalars, but we can still use gauge invariance to gauge away $15$ of them and diagonalize the $6 \times 6$ symmetric traceless matrix. 
Under these assumptions, the scalar sector is finally specified by $5$ degrees of freedom 
which are collected into a single diagonal matrix $\Lambda$:
\[ \Lambda = \left( \begin{array}{cccccc}
 \lambda_1 & & & & & \\
 & \lambda_2 & & & & \\
 & & \lambda_3 & & & \\ 
 & & & \lambda_4 & & \\
 & & & & \lambda_5 & \\
 & & & & & - \sum_{i} \lambda_i
 \end{array} \right).\] 
For generic values of $\lambda_i$ the gauge symmetry is completely broken on the vacuum, 
but there are special configurations such that part of the symmetry is restored. 
In particular in the following sections we will be interested in vacua that break the gauge symmetry to $SO(3)\times SO(3)$ or $SO(4)\times SO(2)$.
Notice that both these groups contain nonabelian factors, 
and thus allow for the possibility to explore solutions with gravity coupled to nonabelian gauge fields, 
which is of main interest here. 
Also notice that both these sunbgroups can be embedded into the compact $SO(6)$ and also in two different noncompact gauge groups, $SO(3,3)$ and $SO(4,2)$ respectively. 
For both these subgroups there is only a single invariant scalar. 
In the $SO(3)\times SO(3)$ case the corresponding matrix is:  
\[ \Lambda_{3,3} = \left( \begin{array}{cccccc}
 \lambda & & & & & \\
 & \lambda & & & & \\
 & & \lambda & & & \\ 
 & & & -\lambda & & \\
 & & & & -\lambda & \\
 & & & & & -\lambda
 \end{array} \right),\]  
\noindent while the $SO(4)\times SO(2)$ invariant scalar is:
\[ \Lambda_{4,2} = \left( \begin{array}{cccccc}
 \lambda & & & & & \\
 & \lambda & & & & \\
 & & \lambda & & & \\ 
 & & & \lambda & & \\
 & & & & -2\lambda & \\
 & & & & & -2\lambda
 \end{array} \right).\]  
\noindent It is now much easier to compute the scalar potentials for a single degree of freedom, as shown in \cite{Gunaydin:1985cu}, 
and the results are the following:
\begin{align} \label{P33P42}
P_{3,3} & = \frac{3g^2}{32}  \left( 6 \sigma + e^{2\lambda} + e^{-2\lambda} \right), \\
P_{4,2} & = \frac{g^2 \sigma}{2}  \left( e^{\lambda} +\frac{\sigma }{2} e^{-2\lambda} \right). 
\end{align}
\noindent where $\sigma = \{+1 , -1 \} $ for the compact and noncompact gaugings respectively 
\footnote{As we will explain later, the potential $P_{4,2}$ also describes a third possible truncation, corresponding to $\sigma=0$.
In this case the scalar potential is vanishing and we get a Mink$_5$ vacuum. 
This choice corresponds to the so called $4^0$ theory in \cite{Romans:1985ps}.}. 

It is easy to see that for $\sigma = +1$, namely for the $SO(6)$ gauging, both the potentials have AdS$_5$ as extremum: 
\begin{equation}
\lambda = 0, \  \  \   P(0)=\frac{3}{4}\ g^2.  
\end{equation}
Notice that this is the only value of $\lambda$ such that the two matrices $\Lambda_{3,3}$ and $\Lambda_{4,2}$ are equal,
while for any other value the solutions in these two sectors are inequivalent. 

For $\sigma = -1$, which describes the $SO(3,3)$ and $SO(4,2)$ gaugings, we get a dS$_5$ extremum and no extrema respectively. 
For the dS vacuum we have:
\begin{equation}
\lambda = 0, \  \  \   P(0)=-\frac{3}{8}\ g^2.  
\end{equation}

\section{$SO(N) \times SO(6-N)$ truncations}
\label{sec:vacua}

\subsection{Overview}

As we already stressed in section \ref{subsec:scalar}, the advantage of looking for solutions in the full maximal supergravity 
lies in the many inequivalent sectors in the space of vacua, corresponding to $H \subset SO(N,6-N)$ invariant truncations
of the maximal gauged theories, or in other words to the possible breaking of the full gauge group to different subgroups. In this paper we will consider $SO(N) \times SO(6-N)$ invariant truncations and give them a lagrangian description. In each truncation there exist black hole like solutions with either abelian or nonabelian gauge fields turned on, together with a large number of product geometries that might correspond to near horizon geometries of static black holes. 

The two most interesting truncations are $SO(4)\times SO(2)$ and $SO(3)\times SO(3)$, 
which can both be embedded in compact and noncompact gauge groups and both contain nonabelian factors.
The choices $SO(4)\times SO(2) \subset SO(6)$ and $SO(4)\times SO(2) \subset SO(4,2)$ correspond the so called $N=4^+$ and $N=4^-$ 
supersymmetric truncations constructed in \cite{Romans:1985ps}. 
On the other hand the $SO(3)\times SO(3)$ invariant truncations do not preserve supersymmetry, 
neither inside the $SO(6)$\cite{Cvetic:2009id} gauging nor inside the $SO(3,3)$ gauging, 
but still can lead to a big variety of interesting non supersymmetric solutions. 
With an abuse of notation we will refer to those truncations as
$3^+$ and $3^-$, where $\sigma=\{+,-\}$ denote the sign of the cosmlogical constant for the corresponding backgrounds\footnote{In our conventions here, AdS has positive, while dS negative cosmological constant.}. 

\begin{table} \begin{center} \label{3} \caption{Main features of $SO(N)\times SO(6-N)$ truncations}
\begin{tabular}{|c|c|c|c|} 
\hline
\begin{minipage}[c][.05\textheight]{.2\textwidth}
\begin{center}
Truncation
\end{center}
\end{minipage}
&
\begin{minipage}[c][.05\textheight]{.3\textwidth}
\begin{center}
Gauge Group
\end{center}
\end{minipage}
&
\begin{minipage}[c][.05\textheight]{.2\textwidth}
\begin{center}
Extremum
\end{center}
\end{minipage}
&
\begin{minipage}[c][.05\textheight]{.2\textwidth}
\begin{center}
SUSY
\end{center}
\end{minipage}
\\
\hline

\begin{minipage}[c][.05\textheight]{.2\textwidth}
\begin{center}
$5^+$
\end{center}
\end{minipage}
&
\begin{minipage}[c][.05\textheight]{.3\textwidth}
\begin{center}
$SO(5)\subset SO(6)$
\end{center}
\end{minipage}
&
\begin{minipage}[c][.05\textheight]{.2\textwidth}
\begin{center}
AdS$_5$
\end{center}
\end{minipage}
&
\begin{minipage}[c][.05\textheight]{.2\textwidth}
\begin{center}
$\times$
\end{center}
\end{minipage}
\\
\hline

\begin{minipage}[c][.05\textheight]{.2\textwidth}
\begin{center}
$5^-$
\end{center}
\end{minipage}
&
\begin{minipage}[c][.05\textheight]{.3\textwidth}
\begin{center}
$SO(5)\subset SO(5,1)$
\end{center}
\end{minipage}
&
\begin{minipage}[c][.05\textheight]{.2\textwidth}
\begin{center}
$\times$
\end{center}
\end{minipage}
&
\begin{minipage}[c][.05\textheight]{.2\textwidth}
\begin{center}
$\times$
\end{center}
\end{minipage}
\\
\hline

\begin{minipage}[c][.05\textheight]{.2\textwidth}
\begin{center}
$4^+$
\end{center}
\end{minipage}
&
\begin{minipage}[c][.05\textheight]{.3\textwidth}
\begin{center}
$SO(4)\times SO(2)\subset SO(6)$
\end{center}
\end{minipage}
&
\begin{minipage}[c][.05\textheight]{.2\textwidth}
\begin{center}
AdS$_5$
\end{center}
\end{minipage}
&
\begin{minipage}[c][.05\textheight]{.2\textwidth}
\begin{center}
$\surd$
\end{center}
\end{minipage}
\\
\hline

\begin{minipage}[c][.05\textheight]{.2\textwidth}
\begin{center}
$4^-$
\end{center}
\end{minipage}
&
\begin{minipage}[c][.05\textheight]{.3\textwidth}
\begin{center}
$SO(4)\times SO(2)\subset SO(4,2)$
\end{center}
\end{minipage}
&
\begin{minipage}[c][.05\textheight]{.2\textwidth}
\begin{center}
$\times$
\end{center}
\end{minipage}
&
\begin{minipage}[c][.05\textheight]{.2\textwidth}
\begin{center}
$\surd$
\end{center}
\end{minipage}
\\
\hline

\begin{minipage}[c][.05\textheight]{.2\textwidth}
\begin{center}
$4^0$
\end{center}
\end{minipage}
&
\begin{minipage}[c][.05\textheight]{.3\textwidth}
\begin{center}
$SO(4)\times SO(2)_g$
\end{center}
\end{minipage}
&
\begin{minipage}[c][.05\textheight]{.2\textwidth}
\begin{center}
Mink$_5$
\end{center}
\end{minipage}
&
\begin{minipage}[c][.05\textheight]{.2\textwidth}
\begin{center}
$\surd$
\end{center}
\end{minipage}
\\
\hline

\begin{minipage}[c][.05\textheight]{.2\textwidth}
\begin{center}
$3^+$
\end{center}
\end{minipage}
&
\begin{minipage}[c][.05\textheight]{.3\textwidth}
\begin{center}
$SO(3)\times SO(3)\subset SO(6)$
\end{center}
\end{minipage}
&
\begin{minipage}[c][.05\textheight]{.2\textwidth}
\begin{center}
AdS$_5$
\end{center}
\end{minipage}
&
\begin{minipage}[c][.05\textheight]{.2\textwidth}
\begin{center}
$\times$
\end{center}
\end{minipage}
\\
\hline

\begin{minipage}[c][.05\textheight]{.2\textwidth}
\begin{center}
$3^-$
\end{center}
\end{minipage}
&
\begin{minipage}[c][.05\textheight]{.3\textwidth}
\begin{center}
$SO(3)\times SO(3)\subset SO(3,3)$
\end{center}
\end{minipage}
&
\begin{minipage}[c][.05\textheight]{.2\textwidth}
\begin{center}
dS$_5$
\end{center}
\end{minipage}
&
\begin{minipage}[c][.05\textheight]{.2\textwidth}
\begin{center}
$\times$
\end{center}
\end{minipage}
\\
\hline

\begin{minipage}[c][.05\textheight]{.2\textwidth}
\begin{center}
$2^+$
\end{center}
\end{minipage}
&
\begin{minipage}[c][.05\textheight]{.3\textwidth}
\begin{center}
$SO(2)^3\subset SO(6)$
\end{center}
\end{minipage}
&
\begin{minipage}[c][.05\textheight]{.2\textwidth}
\begin{center}
AdS$_5$
\end{center}
\end{minipage}
&
\begin{minipage}[c][.05\textheight]{.2\textwidth}
\begin{center}
$\surd$
\end{center}
\end{minipage}
\\
\hline

\begin{minipage}[c][.05\textheight]{.2\textwidth}
\begin{center}
$2^-$
\end{center}
\end{minipage}
&
\begin{minipage}[c][.05\textheight]{.3\textwidth}
\begin{center}
$SO(2)^3\subset SO(4,2)$
\end{center}
\end{minipage}
&
\begin{minipage}[c][.05\textheight]{.2\textwidth}
\begin{center}
$\times$
\end{center}
\end{minipage}
&
\begin{minipage}[c][.05\textheight]{.2\textwidth}
\begin{center}
$\surd$
\end{center}
\end{minipage}
\\
\hline

\end{tabular}
\end{center}
\end{table}

We will not consider the $2^+$ and $2^-$ truncations, corresponding to $SO(2)^3$ invariant sectors inside $SO(6)$ or $SO(4,2)$,
since they only allow for abelian solutions. The $2^+$ corresponds to the $N=2$ supersymmetric truncation formulated in \cite{Cvetic:1999xp}, whose possible solutions were already explored in detail and are of the same type of the abelian solutions in the $4^+$ theory. Similarly the abelian solutions in the $2^-$ truncation correspond to those of the $4^-$.  
We also do not consider the $5^+$ and $5^-$ truncations, which would lead to the same type of solutions as $4^+$ and $4^-$ but without the possibility of preserving supersymmetry.

Finally we will also consider the so called $4^0$ theory, another possible supersymmetric truncation formulated in \cite{Romans:1985ps},
which corresponds to gauging only $SO(2) \subset SO(4)\times SO(2)$. It can be thought of as truncation of a non-maximal gauging of $\nn=8$ that can be realized by the embedding tensor formalism \cite{deWit:2004nw}.  
This gauging has a vanishing scalar potential, and consequently a Mink$_5$ (non-BPS) vacuum. The $4^0$ theory is therefore similar to the 4d $U(1)$ supergravities with vanishing scalar potential \cite{Hristov:2012nu} - it has an interesting BPS spectrum, which we explore in the next section.

The main features of all the possible $SO(N)\times SO(6-N)$ invariant truncations are summarized in table $3$.

\subsection{Effective lagrangian and equations of motion}
\label{subsec:efflag}

We now give an effective lagrangian description to the $SO(N)\times SO(6-N)$ invariant truncations of maximal gauged supergravity.
For our purposes it is enough to consider only the bosonic part, neglecting the tensor fields $B$
which can never be turned on if we look for fully $SO(N)\times SO(6-N)$ invariant vacua.
The strategy is simple - we start from the full lagrangian in \cite{Gunaydin:1985cu} 
and set to zero all the fields that are not invariant under the action of the given subgroup, 
which leads to the following:
\begin{itemize}
 \item We keep one single invariant scalar $\lambda$ and compute the corresponding potential as explained in section \ref{subsec:scalar} and corresponding kinetic terms. 
 \item We set to zero all the tensor fields $B$, which are not invariant under the desired group.
 \item We turn on only the gauge fields that correspond to the generators of $SO(N)\times SO(6-N)$ and compute the corresponding kinetic terms.
 \item We set all the fermions to zero.
 \end{itemize}
\noindent These assumptions simplify dramatically the terms entering the full lagrangian, 
and after some work we get the following bosonic effective lagrangian:
\begin{equation} \label{efflag}
\mathcal{L}= -\frac{1}{4} R + \frac{3 n}{8} \partial_{\mu} \lambda \partial^{\mu} \lambda - P(\lambda)
- \frac{1}{4 n} e^{2\lambda} F_{\mu \nu}^I  F^{I \mu \nu} -\frac{1}{4}e^{-2 n \lambda} \tilde{F}_{\mu \nu}^I \tilde{F}^{I\mu \nu } ,
\end{equation}
\noindent where we defined: 
\begin{equation}
n \equiv \frac{N}{6-N}.
\end{equation}
\noindent The scalar potentials are given in \eqref{P33P42}, 
and the field strenght $F$ and $\tilde{F}$ correspond to the $N$ and $6-N$ gauge directions respectively.

In the case of $SO(4)\times SO(2)$ we have defined:
\begin{align} \label{GF42}
 A^I &=\frac{1}{2} \epsilon^{IJK}A_{JK} + A^{I4}, \\ \nonumber
 \tilde{A}^1 & =A^{56},
\end{align}
\noindent where $I = \{1,2,3\}$. The Yang Mills equations give the constraint: $\frac{1}{2} \epsilon^{IJK}A_{JK} = A^{I4}$,
which means that we only have $SU(2)_L \subset SO(4)$ gauge fields turned on, while the orthogonal $SU(2)_R$ gauge fields are vanishing.

In the case of $SO(3)\times SO(3)$ the gauge fields are: 
\begin{align} \label{GF33}
 A^I &=\frac{1}{2} \epsilon^{IJK}A_{JK}, \\ \nonumber
 \tilde{A}^I &=\frac{1}{2} \epsilon^{I\tilde{J}\tilde{K}}A_{\tilde{J}\tilde{K}},
\end{align}
where $I,\tilde{I} = \{1,2,3\}$. We further have the constraint $F^I \wedge \tilde{F}^J = 0.$
 
Note that in principle we should add the Chern-Simons terms to the lagrangian, which are vanishing in the $SO(3)\times SO(3)$ case,
and proportional to $F \wedge F \wedge \tilde{a}$ for $SO(4) \times SO(2)$. 
However we will ignore these terms since they are completely irrelevant for the type of solutions we will consider, 
as they can only be nonvanishing in the presence of both electric and magnetic fields together.

Starting from the lagrangian \eqref{efflag} we can derive the following equations of motion for bosonic backgrounds:
\begin{align} \label{eom}
0 & = R_{\mu \nu} - \frac{3n}{4} \partial_{\mu} \lambda \partial_{\nu} \lambda +\frac{4}{3} g_{\mu \nu} P + \frac{e^{2\lambda}}{n} \left( 2 F_{\mu \rho }^I F_{\nu}^{I \rho} - \frac{1}{3} g_{\mu \nu}F^2  \right) + e^{-2 n \lambda} \left( 2 \tilde{F}_{\mu \rho }^I \tilde{F}_{\nu}^{I \rho} - \frac{1}{3} g_{\mu \nu}\tilde{F}^2  \right) , \nonumber \\ 
0 & = \frac{n}{2} \Box \lambda +\frac{2}{3} \frac{\partial P}{\partial \lambda} +\frac{1}{3n}e^{2\lambda}F^2 - \frac{n}{3}e^{-2 n \lambda}\tilde{F}^2, \nonumber \\
0 & = D_{\nu}\left( e^{2\lambda} F^{I \nu \mu} \right) = D_{\nu}\left( e^{-2n \lambda} \tilde{F}^{I \nu \mu} \right). 
\end{align} 
In the following sections we will be able to solve these equations in a large number of cases.

\section{Attractor geometries}
\label{sec:exact}
We look for solutions to the equations of motion \eqref{eom} in the case of product geometries $M_d \times \Sigma ^{5 - d}$, 
with constant scalars and gauge fields
\footnote{ For completeness there also exist product geometry solutions $M \times \Sigma$ with vanishing gauge fields and vanishing scalar,
corresponding to the case in which the internal and external manifolds have the same curvature, and the product is thus a five dimensional einstein space 
satisfying $R_{\mu \nu} = \alpha \ g_{\mu \nu} $.
\\ \indent In the $4^+$ and $3^+$ theories we can have both AdS$_3 \times \mathbb{H}^2$ and AdS$_2 \times \mathbb{H}^3$ with $\alpha = g^2$,
while in the $3^-$ theory we can have either dS$_3 \times S^2$ or dS$_2 \times S^3$ with $\alpha = - g^2/2$.
\\ \indent In these cases the residual symmetry on the vacuum is enhanced to the full gauge group: $SO(6)$ or $SO(3,3)$ respectively.
}.
These type of solutions are very interesting as they correspond to near horizon geometries of black holes and black strings. 
\\ In fact in principle any such attractor can be connected with a full black hole like solution to a suitable maximally symmetric vacuum at infinity
\footnote{This is always true except for the attractor geometries in the $4^-$ theory, 
whose potential doesn't admit any maximally symmetric five manifold as extremum, as explained in \ref{subsec:scalar}}.
However, as we will explain in section \ref{sec:analytic}, full analytic solutions can only be found for a particular subclass of the listed attractors,
due to the nontrivial form of the scalar potentials. Another interesting sublcass is given by supersymmetric attractors,
which can be easily connected to a proper asymptotics with a numerical solution to the BPS equations. 

\subsection{Fields ansatz}
\label{subsec:Ansatz}
\subsubsection{Metric ansatz}
We are interested in product geometries M$_{d}^K \times \Sigma ^{5 - d}_k$, 
where $K$ and $k$ label the sign of the curvature of the external and internal space respectively.
We choose the external space to be a maximally symmetric lorentzian $d$-manifold: M$_{d}^K = \{$dS$_d$, Mink$_d$, AdS$_d\}$,
corresponding to $K= \{ 1,0,-1 \}$. We express the metric in global coordinates as:
\begin{equation}
 ds^2_{d,K} = \left( 1 - K \frac{r^2}{L^2} \right) dt^2 - \left( 1 - K \frac{r^2}{L^2} \right)^{-1} dr^2 - r^2 d\Omega^2_{d-2},
\end{equation}
where the parameter $L$ is the length of the external space.

We allow the internal space to be any maximally symmetric euclidean $(5-d)$-manifold: 
$\Sigma^{5-d}_k = \{ S^{5-d}, \mathbb{R}^{5-d} , \mathbb{H}^{5-d} \}$, 
corresponding to $k= \{ 1,0,-1 \}$, with metric:
\begin{equation} \label{intmet}
 ds^2_{5-d,k} =  R^2 \left( d\psi ^2 + f_k(\psi)^2 d\Omega^2_{4-d} \right),
\end{equation}
where the parameter $R$ is the radius and:
\begin{equation}
 f_k(\psi) =
\left\{
\begin{array}{rl}
\sin \psi & \mbox{if } k = 1 \\
\psi & \mbox{if } k = 0 \\
\sinh \psi & \mbox{if } k = - 1
\end{array}
\right.
\end{equation}
In the case of $\mathbb{R}$ and $\mathbb{H}$ we can also compactify the internal space with a suitable quotient,
which is described locally by the same metric
\footnote{Notice that in the case of $k=0$ the metric \eqref{intmet} describes $\mathbb{R}^{5-d}$ is spherical coordinates, 
which is appropriate to describe a black brane near horizon. 
If we wanted to describe a toroidal black hole or a toroidal black string the internal space would be $T^{5-d}$,
in which case we need to use cartesian coordinates with periodic boundary conditions.}.
Geometries of the type $M_2 \times \Sigma^3$ are appropriate to describe Black Holes attractors, 
while $M_3 \times \Sigma^2$ backgrounds correspond to Black Strings.

\subsubsection{Gauge field ansatz}
Given the ansatz for the metric there is a natural guess for a magnetic gauge field, 
which consists on setting it proportional to the spin connection $\omega$ on the internal space: 
\begin{equation} \label{twist}
A = p\  \omega |_{\Sigma},
\end{equation}
where $p$ is the magnetic charge\footnote{More precisely we choose the gauge field to be:  
\begin{equation} \label{twistk}
k A = - p\  \omega |_{\Sigma_k},
\end{equation}
where $k$ is the sign of the curvature on the internal space.
Notice that for $k=0$ the spin connection is pure gauge, so we can set it to zero with a proper gauge transformation. 
Then equation \eqref{twistk} does not fix the gauge field $A$ anymore.
In the case of $\mathbb{R}^2$ it is possible to put a magnetic field strength, whose form is given in \eqref{AFS}, 
while for $\mathbb{R}^3$ we get a constraint on the Yang Mills charge to be zero, see \eqref{NAQ}. The same works for $T^2$ and $T^3$.}. 
This ansatz comes from the so called $twisting$ procedure, which allows to solve the Killing spinor variation in the BPS equations
\footnote{There exists in fact a standard way of solving Killing spinor equations in supersymmetric theories with extended $R$ symmetry, 
which consists on choosing a constant spinor and setting its variation under local Lorentz transformations to be equal to its variation under $R$ symmetry.
In other words the spinor behaves as a scalar. This procedure was first introduced in \cite{Witten:1988ze}, 
and later applied to black holes in AdS space in \cite{Maldacena:2000mw}.}.

In the case of electric field we have to remember that the Coulomb's law in five dimensions determines an $r^{-2}$ scaling on the potential:
\begin{equation} \label{coulomb}
A_t = -\frac{Q}{r^2}. 
\end{equation}
This is all we need to know to determine the field strengths in the following three cases:

\begin{itemize}

 \item \textbf{Nonabelian Magnetic Field:} 
In order to support a nonabelian gauge field an internal space of dimension three is needed, since we want an $SO(3)$ spin connection.
Solutions with nonabelian gauge field can thus describe black holes with various topologies.
We can solve the twisting condition \eqref{twist} explicitely using the metric ansatz \eqref{intmet}, 
and we get the following form for the field strength:
\begin{equation} \label{NAFS}
   F^I = p\ \varepsilon^{IJ L} e^{JL},
\end{equation} 
where $\varepsilon^{I J L}$ are the $SO(3)$ structure constants and  $e^I$ are the vielbein on the internal space. 
The Yang Mills charge is quantized:
\begin{equation} \label{NAQ}
 g\ p = - k. 
\end{equation}
Notice that the nonabelian charge vanishes for $k=0$,
namely a flat three manifld does not support a nonabelian field strength.

\indent We now can ask which amount of gauge symmetry is preserved by the ansatz \ref{NAFS}.
\\In the $4^{\sigma}$ truncations we can only turn on a single nonabelian gauge field, say $SU(2)_L$,
while the abelian gauge field corresponding to the extra $SO(2)$ vanishes.
The residual gauge symmetry is thus $SU(2)_L \times SO(2)$. 
In the case of $3^{\sigma}$ truncations we can instead have two nonvanishing Yang Mills gauge fields, 
which turn to be equal since both $p$ and $\tilde{p}$ are quantized in the same way. This leads to a full $SO(3) \times SO(3)$ invariance.

\item \textbf{Abelian Magnetic Field:} 
It is the type of gauge field which is needed to describe a Black String, 
as it can be supported by an internal space of dimension two.
In this case the field strength turns out to be proportional to the volume form: 
\begin{align} \label{AFS}
   F^1 =  p\ vol_{\Sigma}, \nonumber \\
   \tilde{F}^1=  \tilde{p}\ vol_{\Sigma},
   \end{align} 
A crucial point is that the abelian charge is not quantized, unlike the nonabelian one. 
Due to this feature the Black String Attractors will come in two parameter families, while the nonabelian attractors are isolated points. 
Also notice that a nonvanishing abelian field strength can be supported also by a flat internal space, unlike the nonabelian case. 

\indent It is important to stress that the abelian anzatz \ref{AFS} partially breaks the gauge symmetry
to $U(1)_L  \times SO(2)$ in the $4^{\sigma}$ truncations,
and to $SO(2)\times SO(2)$ in the $3^{\sigma}$ truncations.

\item \textbf{Electric Field:}   
In this case it is natural to choose a three dimensional internal space,
which in five dimensions corresponds to a black hole horizon.
From the Coulomb's law \eqref{coulomb} we derive the electric field strength to be: 
\begin{align}  \label{EFS}
   F_{r t} = Q/r^3, \nonumber \\
   \tilde{F}_{r t} = \tilde{Q}/r^3,
   \end{align} 
where the abelian electric charge $Q$ is also not quantized, and again the gauge symmetry is partially broken. 
If we look for attractor geometries we can replace the radial coordinate $r$ with the horizon radius $R$ in the ansatz.

\end{itemize}

%

\subsection{Attractor solutions} \label{subsec:solutions}
We now plug our ansatz for attractor solutions in the equations of motion \eqref{eom}. 
The spacetime symmetries require the scalar $\lambda$ to be a constant:
\begin{equation}
 e^{\lambda} \equiv c .
\end{equation}
The equations of motion then reduce to three algebraic equations: one scalar equation of motion
plus two Einstein equations, one for the internal and one for the external directions. 
For nonabelian attractors there are three variables: the horizon radius $R$, the length of the external space $L$ and the scalar $c$, 
while the Yang Mills charges $\tilde{p}$ and $p$ are fixed by the quantization condition. 
In the case of abelian attractors the charges are not quantized and the equations depend on all five variables, 
so that attractor geometries come in two parameter families. 
\\ \indent We now list all possible solutions, organized in nonabelian attractors, black string attractors and black hole attractors,
accordingly to the type of gauge field which is turned on.

\subsubsection{Nonabelian attractors}
As we already stressed, the nonabelian charges are necessarily quantized \eqref{NAQ}, 
which implies that there are no free parameters in the equations and the solutions are isolated points. 
Also, it turns out that nonabelian attractors are only possible for compact gaugings.
The results are summarized in the table $4$.

As we shall see, in the $4^+$ theory the AdS$_2 \times \mathbb{H}^3$ attractor is also supersymmetric, 
and the corresponding values for the background parameters are:
\begin{equation} \label{NA42sol}
 g L = 2^{-1/3},\ \ g R = 2^{2/3},\ \ c = 2^{2/3}.
\end{equation}
The attractors in the $3^+$ theory are instead non supersymmetric, and the near horizon values of the parameters are:
\begin{equation} \label{NA33sol}
 2(g L)^2 = 1 - k\left( 4+k^2 \right)^{-1/2},\ \ (g R)^2 = -k + \left( 4+k^2 \right)^{1/2} ,\ \ c = 1.
\end{equation}

\begin{table} \label{4} \begin{center} \caption{Nonabelian Attractors}
\begin{tabular}{|c|c|c|} 
\hline
\begin{minipage}[c][.05\textheight]{.15\textwidth}
\begin{center}

\end{center}
\end{minipage}
&
\begin{minipage}[c][.05\textheight]{.1\textwidth}
\begin{center}
$\textbf{4}^+$
\end{center}
\end{minipage}
&
\begin{minipage}[c][.05\textheight]{.1\textwidth}
\begin{center}
$\textbf{3}^+$
\end{center}
\end{minipage}
\\
\hline

\begin{minipage}[c][.05\textheight]{.15\textwidth}
\begin{center}
AdS$_2 \times S^3$
\end{center}
\end{minipage}
&
\begin{minipage}[c][.05\textheight]{.1\textwidth}
\begin{center}

\end{center}
\end{minipage}
&

\begin{minipage}[c][.05\textheight]{.1\textwidth}
\begin{center}
$\surd$
\end{center}
\end{minipage}
\\
\hline

\begin{minipage}[c][.05\textheight]{.15\textwidth}
\begin{center}
AdS$_2 \times \mathbb{H}^3$
\end{center}
\end{minipage}
&
\begin{minipage}[c][.05\textheight]{.1\textwidth}
\begin{center}
$\surd$
\end{center}
\end{minipage}
&
\begin{minipage}[c][.05\textheight]{.1\textwidth}
\begin{center}
$\surd$
\end{center}
\end{minipage}
\\
\hline

\end{tabular}
\end{center}
\end{table}

\subsubsection{Black string attractors}

Abelian solutions come in two parameter families, since the gauge coupling constants are not quantized.
We choose to express $(p, \tilde{p}, L)$ as functions of $(R,c)$ which are kept as free parameters.
We are then able to express all the possible black string attractor geometries in a compact form.
For the $4^{\sigma}$ truncations the solutions are given by:
\begin{align} \label{42Asol}
 & p^2 = R^2 c^{-2} \left( k + g^2 R^2 c \sigma \right)  , \nonumber \\
 & \tilde{p}^2 = \frac{R^2 c^{4}}{4} \left( k + g^2 R^2 c^{-2} \sigma^2 \right), \nonumber \\
 & \frac{-K }{L^2} = \frac{k}{4 R^2} + \frac{g^2 \sigma (2c + \sigma c^{-2})}{4},
 \end{align}
where the quantities $(K,k)$ were defined in section \ref{subsec:Ansatz} to be the sign of the curvatures of the external and internal space respectively,
and $\sigma=(+1,-1,0)$ for the $(4^+,4^-,4^0)$ gaugings respectively. 
In the $3^{\sigma}$ truncations we get:
\begin{align} \label{33Asol}
 & p^2 =\frac{3}{64} R^2 c^{-2} \left( 8 k + g^2 R^2 (3c^2-c^{-2}+6\sigma) \right)  , \nonumber \\
 & \tilde{p}^2 =\frac{3}{64} R^2 c^{2} \left( 8 k + g^2 R^2 (3c^{-2}-c^{2}+6\sigma) \right), \nonumber \\
 & \frac{-K }{L^2} = \frac{k}{4 R^2} + \frac{3 g^2 (c^2 + c^{-2}+6\sigma)}{32}.
 \end{align}
The admissible solutions to the systems \eqref{42Asol} and \eqref{33Asol} are summarized in table \ref{5}. 
Notice that we have a large variety of possible attractor geometries, 
and for each geometry we have a two dimensional moduli space \footnote{
This is always true except for the case of flat external space: when $K=0$ the external length $L$ is not defined, 
and in fact it disappears from the equations. The third equation then allows to express $R$ as a function of $c$,
which remains the only free parameter.}.  
As we already mentioned there are two interesting subsets of solutions in the moduli space, which we will analyze later on:
those that can be connected to infinity with a full analytic solution and those that are BPS.

\begin{table} \begin{center} \label{5} \caption{Black String Attractors}
\begin{tabular}{|c|c|c|c|c|c|} 
\hline
\begin{minipage}[c][.05\textheight]{.15\textwidth}
\begin{center}

\end{center}
\end{minipage}
&
\begin{minipage}[c][.05\textheight]{.1\textwidth}
\begin{center}
$\textbf{4}^+$
\end{center}
\end{minipage}
&
\begin{minipage}[c][.05\textheight]{.1\textwidth}
\begin{center}
$\textbf{4}^-$
\end{center}
\end{minipage}
&
\begin{minipage}[c][.05\textheight]{.1\textwidth}
\begin{center}
$\textbf{4}^0$
\end{center}
\end{minipage}
&
\begin{minipage}[c][.05\textheight]{.1\textwidth}
\begin{center}
$\textbf{3}^+$
\end{center}
\end{minipage}

&
\begin{minipage}[c][.05\textheight]{.1\textwidth}
\begin{center}
$\textbf{3}^-$
\end{center}
\end{minipage}
\\
\hline

\begin{minipage}[c][.05\textheight]{.15\textwidth}
\begin{center}
AdS$_3 \times S^{2}$
\end{center}
\end{minipage}
&
\begin{minipage}[c][.05\textheight]{.1\textwidth}
\begin{center}
$\surd$
\end{center}
\end{minipage}
&
\begin{minipage}[c][.05\textheight]{.1\textwidth}
\begin{center}
$\surd$
\end{center}
\end{minipage}
&
\begin{minipage}[c][.05\textheight]{.1\textwidth}
\begin{center}
$\surd$
\end{center}
\end{minipage}
&
\begin{minipage}[c][.05\textheight]{.1\textwidth}
\begin{center}
$\surd$
\end{center}
\end{minipage}
&
\begin{minipage}[c][.05\textheight]{.1\textwidth}
\begin{center}
$\surd$
\end{center}
\end{minipage}
\\
\hline

\begin{minipage}[c][.05\textheight]{.15\textwidth}
\begin{center}
AdS$_3 \times \mathbb{R}^{2}$
\end{center}
\end{minipage}
&
\begin{minipage}[c][.05\textheight]{.1\textwidth}
\begin{center}
$\surd$
\end{center}
\end{minipage}
&
\begin{minipage}[c][.05\textheight]{.1\textwidth}
\begin{center}

\end{center}
\end{minipage}
&
\begin{minipage}[c][.05\textheight]{.1\textwidth}
\begin{center}

\end{center}
\end{minipage}
&
\begin{minipage}[c][.05\textheight]{.1\textwidth}
\begin{center}
$\surd$
\end{center}
\end{minipage}
&\begin{minipage}[c][.05\textheight]{.1\textwidth}
\begin{center}

\end{center}
\end{minipage}

\\
\hline

\begin{minipage}[c][.05\textheight]{.15\textwidth}
\begin{center}
AdS$_3 \times \mathbb{H}^{2}$
\end{center}
\end{minipage}
&
\begin{minipage}[c][.05\textheight]{.1\textwidth}
\begin{center}
$\surd$
\end{center}
\end{minipage}
&
\begin{minipage}[c][.05\textheight]{.1\textwidth}
\begin{center}

\end{center}
\end{minipage}
&
\begin{minipage}[c][.05\textheight]{.1\textwidth}
\begin{center}

\end{center}
\end{minipage}
&
\begin{minipage}[c][.05\textheight]{.1\textwidth}
\begin{center}
$\surd$
\end{center}
\end{minipage}
&\begin{minipage}[c][.05\textheight]{.1\textwidth}
\begin{center}

\end{center}
\end{minipage}

\\
\hline

\begin{minipage}[c][.05\textheight]{.15\textwidth}
\begin{center}
Mink$_3 \times S^{2}$
\end{center}
\end{minipage}
&
\begin{minipage}[c][.05\textheight]{.1\textwidth}
\begin{center}

\end{center}
\end{minipage}
&
\begin{minipage}[c][.05\textheight]{.1\textwidth}
\begin{center}
$\surd$
\end{center}
\end{minipage}
&
\begin{minipage}[c][.05\textheight]{.1\textwidth}
\begin{center}

\end{center}
\end{minipage}
&
\begin{minipage}[c][.05\textheight]{.1\textwidth}
\begin{center}

\end{center}
\end{minipage}
&\begin{minipage}[c][.05\textheight]{.1\textwidth}
\begin{center}
$\surd$
\end{center}
\end{minipage}

\\
\hline

\begin{minipage}[c][.05\textheight]{.2\textwidth}
\begin{center}
dS$_3 \times S^{2}$
\end{center}
\end{minipage}
&
\begin{minipage}[c][.05\textheight]{.1\textwidth}
\begin{center}

\end{center}
\end{minipage}
&
\begin{minipage}[c][.05\textheight]{.1\textwidth}
\begin{center}
$\surd$
\end{center}
\end{minipage}
&
\begin{minipage}[c][.05\textheight]{.1\textwidth}
\begin{center}

\end{center}
\end{minipage}
&
\begin{minipage}[c][.05\textheight]{.1\textwidth}
\begin{center}

\end{center}
\end{minipage}
&\begin{minipage}[c][.05\textheight]{.1\textwidth}
\begin{center}
$\surd$
\end{center}
\end{minipage}

\\
\hline
\end{tabular}
\end{center}
\end{table}

\subsubsection{Black hole attractors}

The solutions in the $4^{\sigma}$ truncations are given by:
\begin{align} \label{42Esol}
 & Q^2 = R^4 c^{-4} \left( 4 c^2 k + \sigma g^2 R^2 (\sigma + c^{-3}) \right), \nonumber \\
 & \tilde{Q}^2 = R^4 c^4   \left( k + \frac{1}{2} \sigma c g^2 R^2 \right), \nonumber \\
 & \frac{-K}{L^2} =  \frac{4k}{R^2} + \sigma g^2\left(2 c +  \sigma c^{-2} \right),
 \end{align}
while for the $3^{\sigma}$ truncations we get:
\begin{align} \label{33Esol}
 & Q^2 = \frac{3}{16} c^{-2} R^4  \left( 8 k + g^2 R^2 ( c^{-2} + 3\sigma ) \right), \nonumber \\
 & \tilde{Q}^2 =  \frac{3}{16} c^2 R^4  \left( 8 k + g^2 R^2 ( c^2 + 3\sigma ) \right), \nonumber \\
 & \frac{-K}{L^2} =  \frac{4k}{R^2} + \frac{ 3 g^2\left( c^2 + c^{-2} + 6\sigma  \right)}{8}.
 \end{align}
The admissible solutions to \eqref{42Esol} and \eqref{33Esol} are summarized in table $6$.

Black hole attractor geometries also come in two dimensional moduli spaces, 
with special subsets corresponding to near horizons of full analytic solutions.
No supersymmetry is ever preserved by this type of solutions.

\begin{table}
\begin{center}
 \label{6} \caption{Black Hole Attractors}
\begin{tabular}{|c|c|c|c|c|c|} 
\hline
\begin{minipage}[c][.05\textheight]{.15\textwidth}
\begin{center}

\end{center}
\end{minipage}
&
\begin{minipage}[c][.05\textheight]{.1\textwidth}
\begin{center}
$\textbf{4}^+$
\end{center}
\end{minipage}
&
\begin{minipage}[c][.05\textheight]{.1\textwidth}
\begin{center}
$\textbf{4}^-$
\end{center}
\end{minipage}
&
\begin{minipage}[c][.05\textheight]{.1\textwidth}
\begin{center}
$\textbf{4}^0$
\end{center}
\end{minipage}
&
\begin{minipage}[c][.05\textheight]{.1\textwidth}
\begin{center}
$\textbf{3}^+$
\end{center}
\end{minipage}

&
\begin{minipage}[c][.05\textheight]{.1\textwidth}
\begin{center}
$\textbf{3}^-$
\end{center}
\end{minipage}
\\
\hline

\begin{minipage}[c][.05\textheight]{.15\textwidth}
\begin{center}
AdS$_2 \times S^{3}$
\end{center}
\end{minipage}
&
\begin{minipage}[c][.05\textheight]{.1\textwidth}
\begin{center}
$\surd$
\end{center}
\end{minipage}
&
\begin{minipage}[c][.05\textheight]{.1\textwidth}
\begin{center}
$\surd$
\end{center}
\end{minipage}
&
\begin{minipage}[c][.05\textheight]{.1\textwidth}
\begin{center}
$\surd$
\end{center}
\end{minipage}
&
\begin{minipage}[c][.05\textheight]{.1\textwidth}
\begin{center}
$\surd$
\end{center}
\end{minipage}
&
\begin{minipage}[c][.05\textheight]{.1\textwidth}
\begin{center}
$\surd$
\end{center}
\end{minipage}
\\
\hline

\begin{minipage}[c][.05\textheight]{.15\textwidth}
\begin{center}
AdS$_2 \times \mathbb{R}^3$
\end{center}
\end{minipage}
&
\begin{minipage}[c][.05\textheight]{.1\textwidth}
\begin{center}
$\surd$
\end{center}
\end{minipage}
&
\begin{minipage}[c][.05\textheight]{.1\textwidth}
\begin{center}

\end{center}
\end{minipage}
&
\begin{minipage}[c][.05\textheight]{.1\textwidth}
\begin{center}

\end{center}
\end{minipage}
&
\begin{minipage}[c][.05\textheight]{.1\textwidth}
\begin{center}
$\surd$
\end{center}
\end{minipage}
&\begin{minipage}[c][.05\textheight]{.1\textwidth}
\begin{center}

\end{center}
\end{minipage}

\\
\hline

\begin{minipage}[c][.05\textheight]{.15\textwidth}
\begin{center}
AdS$_2 \times \mathbb{H}^{3}$
\end{center}
\end{minipage}
&
\begin{minipage}[c][.05\textheight]{.1\textwidth}
\begin{center}
$\surd$
\end{center}
\end{minipage}
&
\begin{minipage}[c][.05\textheight]{.1\textwidth}
\begin{center}

\end{center}
\end{minipage}
&
\begin{minipage}[c][.05\textheight]{.1\textwidth}
\begin{center}

\end{center}
\end{minipage}
&
\begin{minipage}[c][.05\textheight]{.1\textwidth}
\begin{center}
$\surd$
\end{center}
\end{minipage}
&\begin{minipage}[c][.05\textheight]{.1\textwidth}
\begin{center}

\end{center}
\end{minipage}

\\
\hline

\begin{minipage}[c][.05\textheight]{.15\textwidth}
\begin{center}
Mink$_2 \times S^{3}$
\end{center}
\end{minipage}
&
\begin{minipage}[c][.05\textheight]{.1\textwidth}
\begin{center}

\end{center}
\end{minipage}
&
\begin{minipage}[c][.05\textheight]{.1\textwidth}
\begin{center}

\end{center}
\end{minipage}
&
\begin{minipage}[c][.05\textheight]{.1\textwidth}
\begin{center}

\end{center}
\end{minipage}
&
\begin{minipage}[c][.05\textheight]{.1\textwidth}
\begin{center}

\end{center}
\end{minipage}
&\begin{minipage}[c][.05\textheight]{.1\textwidth}
\begin{center}
$\surd$
\end{center}
\end{minipage}

\\
\hline

\begin{minipage}[c][.05\textheight]{.15\textwidth}
\begin{center}
dS$_2 \times S^{3}$
\end{center}
\end{minipage}
&
\begin{minipage}[c][.05\textheight]{.1\textwidth}
\begin{center}

\end{center}
\end{minipage}
&
\begin{minipage}[c][.05\textheight]{.1\textwidth}
\begin{center}

\end{center}
\end{minipage}
&
\begin{minipage}[c][.05\textheight]{.1\textwidth}
\begin{center}

\end{center}
\end{minipage}
&
\begin{minipage}[c][.05\textheight]{.1\textwidth}
\begin{center}

\end{center}
\end{minipage}
&\begin{minipage}[c][.05\textheight]{.1\textwidth}
\begin{center}
$\surd$
\end{center}
\end{minipage}

\\
\hline
\end{tabular}
\end{center}
\end{table}

\section{Supersymmetric attractors and RG flows}
\label{sec:RGflows}
In this section we want to highlight an important region in the moduli space of attractor geometries, the subset of BPS attractors.
These are in fact the most interesting type of near horizons in the context of holographic RG flows and black hole entropy counting, since supersymmetry allows for precision tests and applications of the AdS/CFT correspondence. Some of the BPS black hole and string solutions that we comment on in subsection \ref{subsec:RG}, together with other similar BPS RG flows between AdS vacua \cite{Girardello:1999bd} have already proven very useful for the understanding of the behavior of supersymmetric field theories. It would be equally interesting to use the dual field theories at the BPS attractors that we present in order to account for the macroscopic entropy of the various black objects.

\subsection{BPS Attractors} 
\label{subsec:BPS}
As we already pointed out, the $3^{\sigma}$ truncations do not preserve supersymmetry, so we can restrict ourselves to $4^{\sigma}$ truncations.
These are equivalent to the $\mathcal{N} = 4$ supersymmetric theories formulated in \cite{Romans:1985ps} which have $SO(5)$ as $R$ symmetry group,
out of which the $SO(3) \times SO(2)$ maximal subgroup is gauged. 

It is much simpler to directly use the BPS variations in the $\mathcal{N}=4$ language
\footnote{The matching between the two theories $\nn=4$ and $\nn=8$ works out after the following redefinitions:
\begin{equation}
  \phi= \sqrt{\frac{3}{2}} \lambda,\ \ g_1 =  g,\ \ g_2 = \sqrt{2} \sigma g, \ \ F^{I(4)} = \frac{1}{\sqrt{2}} F^{I(8)} .
\end{equation}
Actually one can start from the BPS equations in maximal gauged supergravity
which are given in \cite{Gunaydin:1985cu}, which however are way more complicated and require computing many non trivial tensors.
After restricting to the $SO(4) \times SO(2)$ invariant sector with the ansatz given in section \ref{subsec:Ansatz} 
and after imposing suitable projections on the Killing spinor, all the gaugino variations $\delta \chi ^{abc}$ collapse to a single equation
which is equivalent to the gaugino equation in \cite{Romans:1985ps}. 
}
which, after setting to zero all the fermions and the tensor fields, reduce to: 
\begin{align} \label{BPSeq}
\delta \psi_{\mu} & = D_{\mu} \epsilon + \frac{g}{12}\left( e^{2\lambda}  + \frac{ \sigma}{2} e^{-\lambda} \right) \Gamma^{45} \epsilon 
- \frac{1}{12} (\gamma_{\mu}^{\nu \rho} - 4\delta_{\mu}^{\nu} \gamma^{\rho})  \left( e^{\lambda} F_{\mu \nu}^I \Gamma^I + e^{-2\lambda} \tilde{F}_{\mu \nu} \right) \epsilon , \\  
 \frac{\delta \chi}{\sqrt{3}} & = \frac{1}{2} \gamma^{\mu}\partial_{\mu} \lambda \epsilon + \frac{g}{6}\left( - e^{2\lambda}  + \sigma e^{-\lambda} \right) \Gamma^{45} \epsilon 
- \frac{1}{12} \gamma^{\mu \nu } \left( e^{\lambda} F_{\mu \nu}^I \Gamma^I -2 e^{-2\lambda} \tilde{F}_{\mu \nu} \right) \epsilon.
\end{align}
\noindent The covariant derivative acts on the Killing spinor as: 
%
%
\begin{equation}
D_{\mu} \epsilon = \nabla_{\mu} \epsilon + 
\frac{1}{2}\ g \tilde{A}_{\mu} \Gamma^{45} \epsilon + \frac{\sigma}{\sqrt{2}}\ g A^I_{\mu} \Gamma^{I45} \epsilon,
\end{equation}
where the gamma matrices $\Gamma^I$ and $\Gamma^{45}$ correspond to the $SO(3)$ and $SO(2)$ gauged subgroups inside $SO(5)$.

After plugging our ansatz for attractor geometries into the BPS variations, these collapse to three independent equations.
We can now identify which subset of the solutions we found in section \ref{sec:vacua} preserve some supersymmetry,
restricting our attention to the case of magnetic backgrounds, since static solutions with electric field turned are never BPS.
We make use of the standard twisting technique to solve the Killing spinor equation $\nabla \epsilon  = 0$ on the internal space,
namely we choose a constant spinor and set its variation under gauge transformation to be equal 
to its variation under local Lorentz transformation. 
In other words we can set the gauge field to be proportional to the spin connection, as defined in \eqref{twistk},
and then impose a suitable projection on the Killing spinor which is needed to identify spacetime indices with gauge indices.

Nonabelian BPS black hole attractors can be obtained after imposing the following two projections: 
%
\begin{equation} \label{NAKSP}
\gamma^{IJ} \epsilon= \frac{1}{2} \varepsilon^{IJK} \Gamma^{K45} \epsilon,
\end{equation}
and the usual quantization condition for the Yang Mills charge \ref{NAQ}. 
The AdS$_2 \times \mathbb{H}^3$ solution of the $4^+$ theory given in \eqref{NA42sol} turns out to be supersymmetric.

The range of possible BPS black string attractors is much broader. 
The following two projections are required
\footnote{We wrote the Killing spinor projections \eqref{NAKSP} and \eqref{AKSP} in the $\mathcal{N}=4$ index notation,
while in the $\mathcal{N}=8$ language we need to impose different projections. In the case of nonabelian attractors the required projections are:
\begin{align} 
\gamma^{IJ} \epsilon = \Gamma^{IJ} \epsilon = \frac{1}{2} \varepsilon^{IJK}\Gamma^{K4} \epsilon, 
\end{align}
while in the abelian case we require:
\begin{align}
\gamma^{23} \epsilon = \Gamma^{23} \epsilon = \Gamma^{14} \epsilon = \Gamma^{56} \epsilon.
\end{align}
Notice that in both the cases in $\nn=8$ theory we need to impose one extra projection, 
which means that these solutions preserve the same amount of supersymmetry in the $\mathcal{N}=8$ and $\mathcal{N}=4$ theories, namely four supercharges.
}:
\begin{align} \label{AKSP}
& \Gamma^{145} \epsilon = \gamma^{23} \epsilon, \nonumber \\
& \Gamma^{3} \epsilon = \epsilon,
\end{align}
together with the quantization condition: 
\begin{equation} \label{AQ}
 g(\sigma p + \tilde{p}) = - k.
\end{equation}




This is the extra condition to be added to the three coming from the BPS variations, 
which selects a one dimensional subspace inside the moduli space of AdS$_3 \times \Sigma^2_k$ black string attractors. 
This subspace is determined by:
\begin{align} \label{42BPS}
 \frac{k}{p} &=  \frac{g(c^3 - 4\sigma)}{2} , \nonumber \\
 \frac{1}{L} &= \frac{g(c^3 +2 \sigma) }{4 c } ,
 \end{align}
where the horizon radius is fixed in terms of the magnetic charge as:
\begin{equation}
 p = \frac{c g R^2}{2}.
\end{equation}
Since we had to impose two projections \ref{AKSP} on the Killing spinor, the amount of preserved supersymmetry is $1/4$, corresponding to 4 supercharges. 

For completeness there also exists a single Mink$_3 \times S^2$ background which is also BPS in the $4^-$ theory for the following values of the parameters:
\begin{equation} \label{M3S2}
g p = \frac{1}{3},\ \ gR= \frac{2^{1/3}}{3^{1/2}},\ \ c=2^{1/3}. 
\end{equation}
However we cannot refer to this background as attractor, 
since the $4^-$ theory doesn't have any maximally symmetric vacuum to interpolate with at infinity.

To conclude this section we want to stress that the black string attractor solutions to the equations of motion \eqref{42Asol} 
reduce to the solutions to the BPS equations after imposing the quantization condition \eqref{AQ}.
Also notice that, being the BPS equations linear in the fields, 
we lost the $\mathbb{Z}_2 \times \mathbb{Z}_2$ symmetry acting on the signs of the magnetic charge,
namely the BPS equations fix the sign of the magnetic charge.

\subsection{Moduli space of black string attractors}
\label{subsec:Moduli}

It is interesting to analyze in detail the moduli space of attractor geometries, 
expecially in the case where there are regions with enhanced supersymmetry.
We consider black string attractor geometries of the type AdS$_3 \times \Sigma^2_k$,
and compare their moduli space in the three inequivalent $\nn=4$ theories, together with the moduli space of the ungauged theory.

We decide to invert the equations \eqref{42Asol} in favour of the magnetic charges,
and plot the moduli space in the $(p, \tilde{p})$ plane, where it looks particularly enlightning.
\begin{itemize}
 \item \textbf{AdS$_3$ $\times$ S$^2$ Attractors} 
 \\ This type of geometry exists in all the three theories, for  different ranges of the parameters.
 In the $4^+$ case the equations of motion can be solved in any point of the plane, except for the the axis $p=0$, $\tilde{p}=0$.
 In other words we necessarily need both the gauge filds to be tuned on.
 The BPS equations can instead be solved when the two magnetic charges have opposite sign, in particular we get the constraint:
 \begin{equation}
  gp > 0,\ \ g\tilde{p} <-1.
 \end{equation}
 In the $4^-$ theory the equations of motion admit a solution for $\tilde{p} \neq 0$,
 so we have the possibility to switch on a single gauge filed. 
 We can still have supersymmetry when the two magnetic charges have opposite sign, 
 but the allowed range for the parameters is much smaller and correspond to the rectangle:
 \begin{equation}
  0 < gp < \frac{1}{3},\ \ -1 < g\tilde{p} < -\frac{2}{3}.
 \end{equation}
 It is somehow expected that supersymmetry be harder to get for a non compact gauging.
 \begin{figure}[H]
\centering
{\includegraphics[height=60mm]{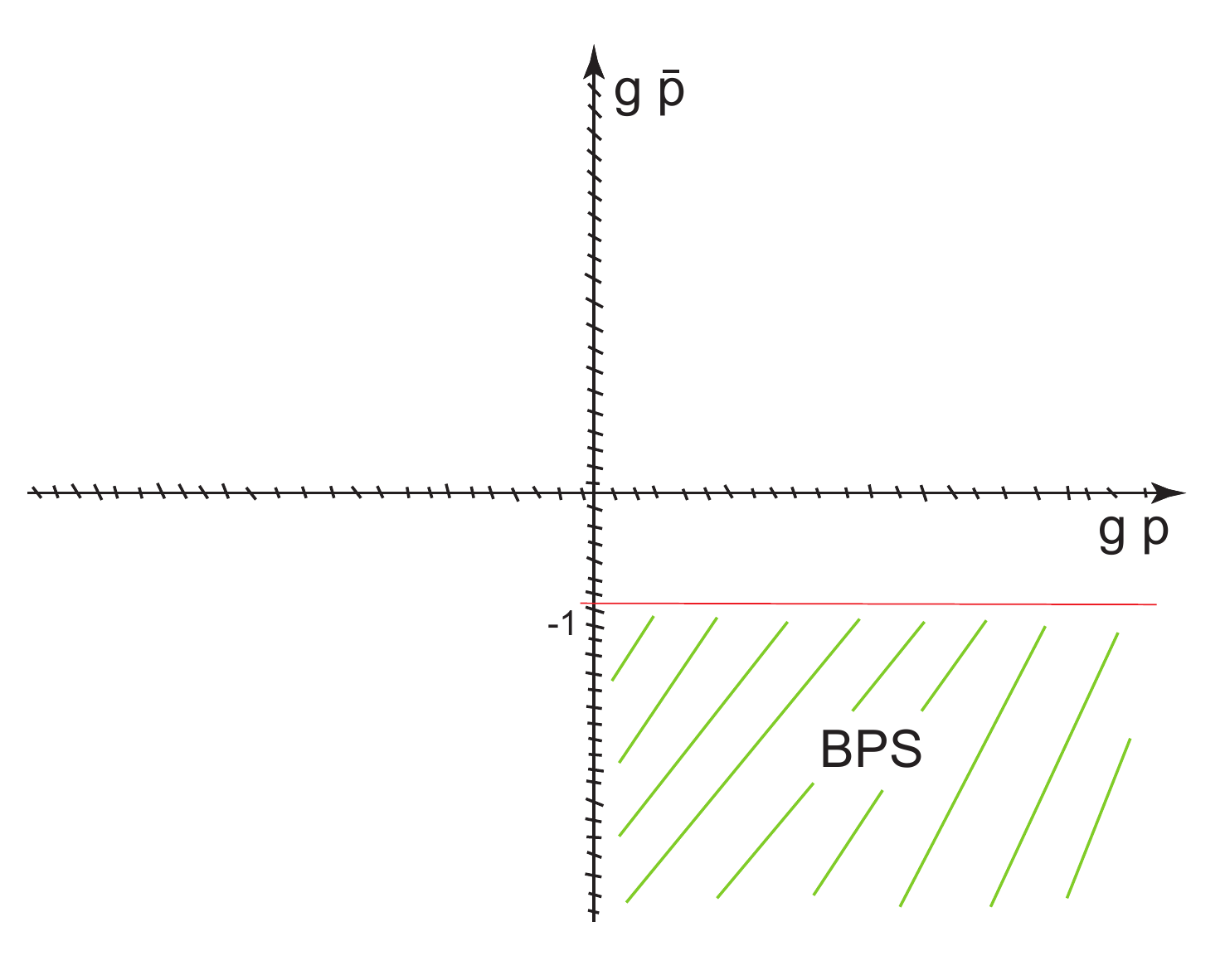}}
{\includegraphics[height=60mm]{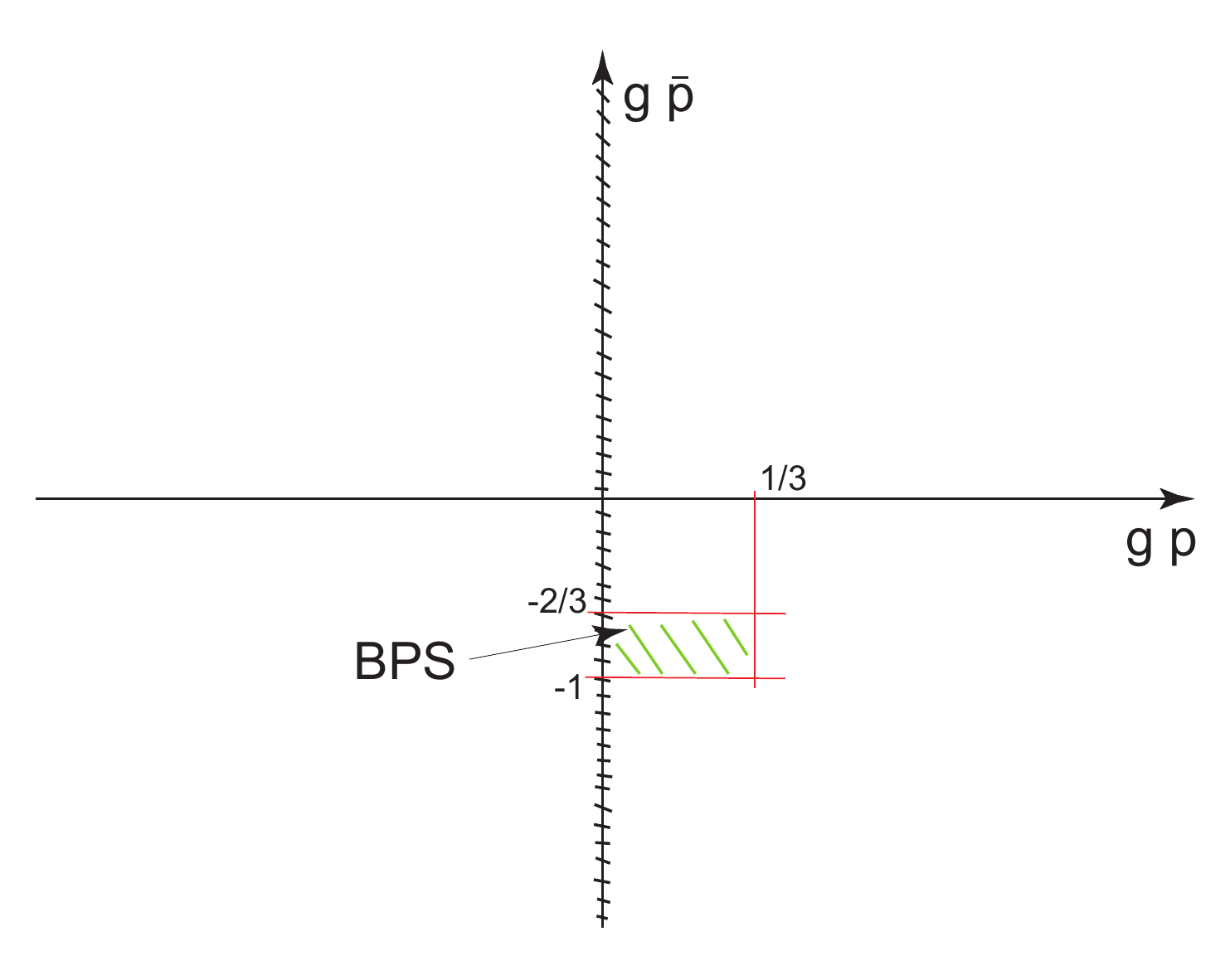}}
\caption{Parameter space for AdS$_3 \times$S$^2$ in the 4$^+$ and 4$^-$ theories.}
\label{fig1}
\end{figure} 
Finally in the case of $4^0$ the allowed range is:
  \begin{equation} \label{p40}
  gp > 0,\ \ g\tilde{p} = -1.
 \end{equation}
 However in this case we have a Minkowski vacuum at infinity, then we can really consider $g$ as a free parameter,
 since it is not needed to fix the length $L_5$. The constraint \eqref{p40} can thus be rewritten as:
 \begin{equation}
  p\ \tilde{p} < 0.
 \end{equation}
 Notice that in this case the equations of motion can be solved if both the gauge fields are turned on,
 namely $p \neq 0$ $\tilde{p} \neq 0$, 
 while the BPS equations select one half of the region where the two magnetic charges have opposite sign.
 Also notice that in the ungauged theory the BPS equations select the complementary region in the moduli space,
 namely there we have the constraint $  p\ \tilde{p} > 0$. This is in exact analogy with the BPS and almost-BPS black hole horizons in 4d ungauged supergravity \cite{Goldstein:2008fq,Bena:2009ev}. Also there the two sectors are related by flipping signs for the charges and the "almost-BPS" horizons are supersymmetric in gauged supergravity with vanishing scalar potential \cite{Hristov:2012nu}.
  \begin{figure}[H]
\centering
{\includegraphics[height=60mm]{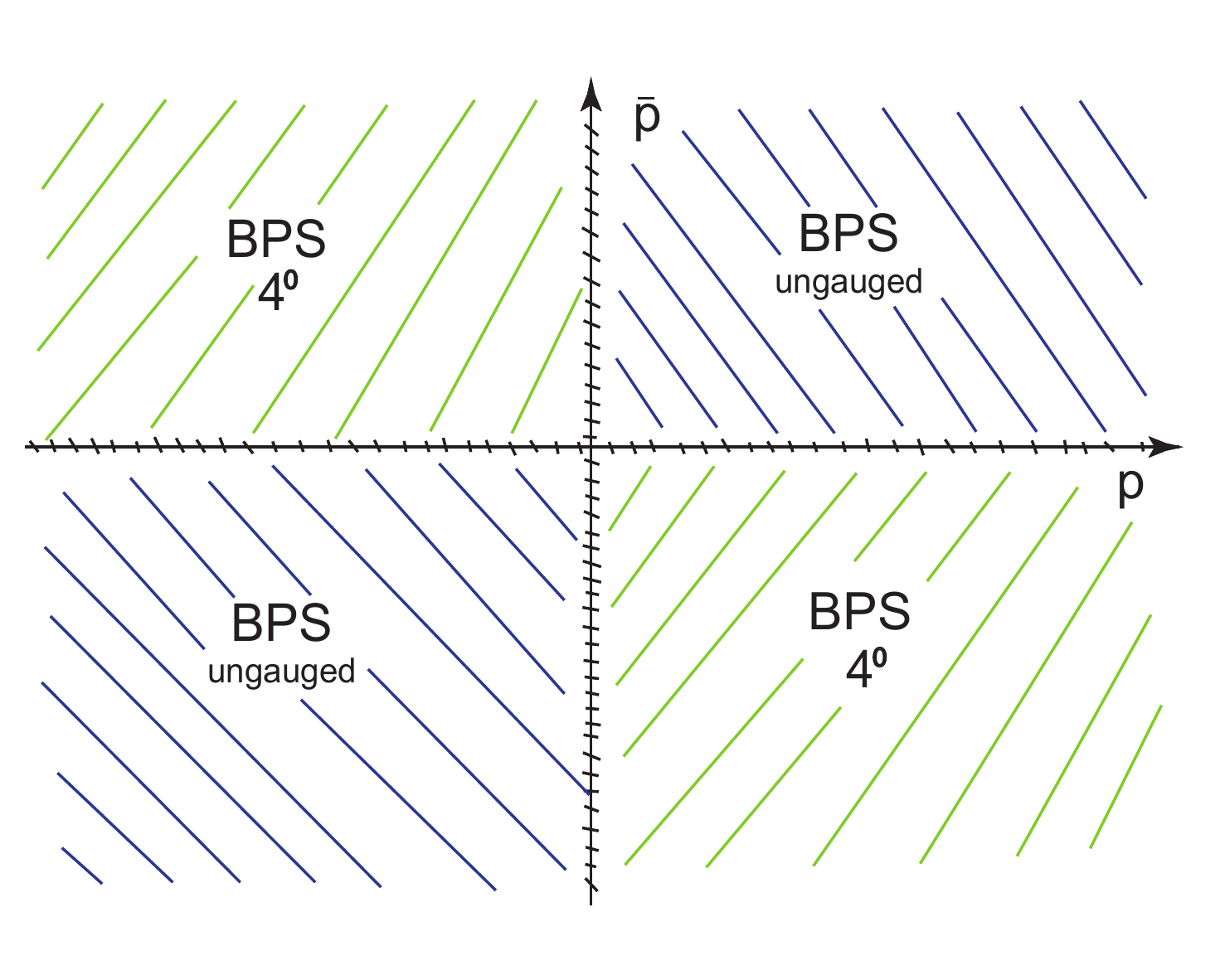}}
\caption{Parameter space for AdS$_3 \times$S$^2$ in the 4$^0$ and ungauged theories.}
\label{fig2}
\end{figure}  
\item \textbf{AdS$_3$ $\times$ $\mathbb{H}^2$ Attractors} 
\\ This type of solution only exist in the $4^+$ theory. The equations of motion can be solved in the whole plane,
while supersymmetry enforces the two cherges to have opposite sign. The BPS region is:
  \begin{equation} \label{p40}
  gp > 0,\ \ g\tilde{p} < 1.
 \end{equation}

 \item \textbf{AdS$_3$ $\times$ $\mathbb{R}^2$ Attractors} 
\\ This type of solution also exist only in the $4^+$ theory. 
The equations of motion can be solved in the region $p\neq 0, \tilde{p}\neq 0$, 
while supersymmetry selects one quarter:
  \begin{equation} \label{p40}
  gp > 0,\ \ g\tilde{p} < 0.
 \end{equation}

 \begin{figure}[H]
\centering
{\includegraphics[height=60mm]{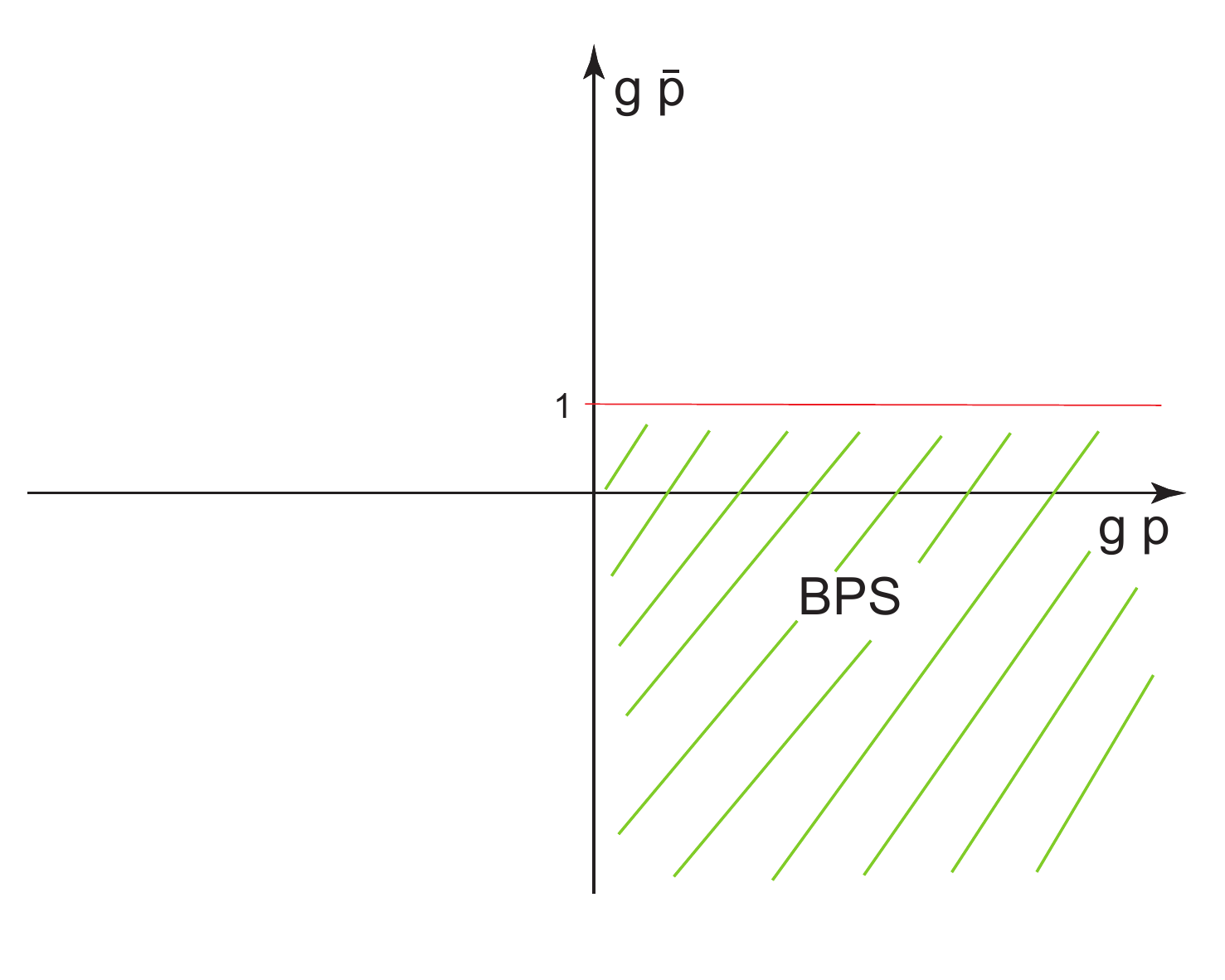}}
{\includegraphics[height=60mm]{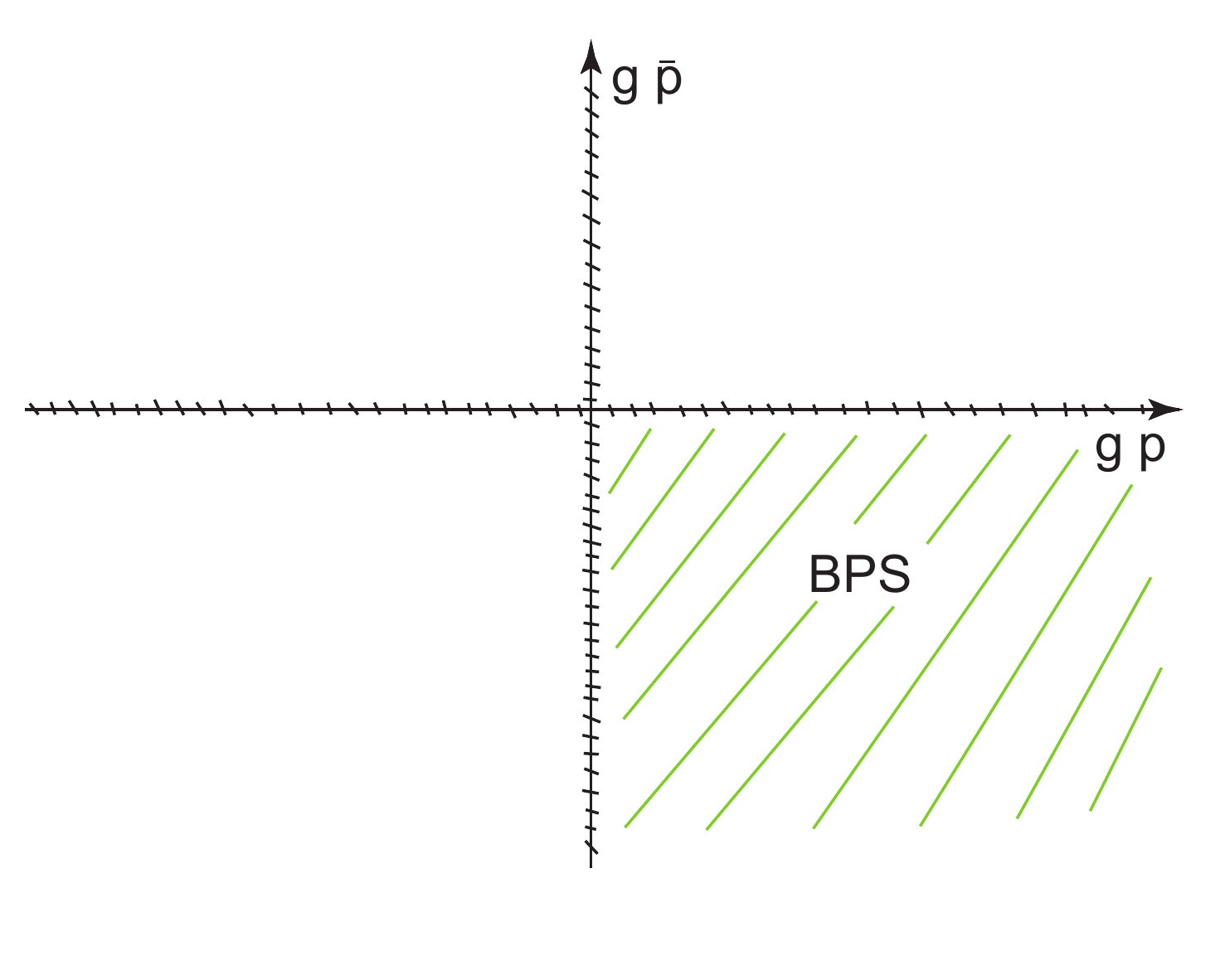}}
\caption{Parameter space for AdS$_3 \times \mathbb{H}^2$ and AdS$_3 \times \mathbb{R}^2$ in the $4^+$ theory}
\label{fig3}
\end{figure}   

 \end{itemize}
 
\subsection{Holographic RG flows}
 \label{subsec:RG}
We already hinted at the possibility of connecting the AdS$_d \times \Sigma^{5-d}$ attractors to AdS$_5$ at infinity 
with a numerical solution of the BPS equations. 
This leads to one of the most interesting applications of black hole in gauged supergravity to AdS/CFT correspondence,
since it suggets the existence of a holographic RG flow between a conformal field theory in four dimensions in the UV and a one or two dimensional one in the IR.

The interpolating solution for the AdS$_2 \times \mathbb{H}^3$ nonabelian attractor has already been constructed in \cite{Nieder:2000kc,Naka:2002jz},
so we focus on black strings.
A proper ansatz for the metric of a full black string solution is the following:
\begin{equation}
  ds^2 = f(r)^2 (dt^2 - dz^2) - f(r)^{-2}dr^2 - h(r)^2 ds^2_{\Sigma},
 \end{equation}
where we allowed all the unknown functions to depend on the radial coordinate only,
and we also allow for running scalars $\lambda (r)$.  
Since we want to interpolate between AdS$_3 \times \Sigma^2$ and AdS$_5$ we have to impose proper boundary conditions,
which at the horizon are:
\begin{equation}
 f=\frac{r}{L} ,\ \ h = R,\ \ e^{\lambda} = c,
\end{equation}
corresponding to AdS$_3$ in Poincare' coordinates with length $L$.
At infinity we impose:
\begin{equation}
 f=\frac{ g r}{2} ,\ \ h = r,\ \ e^{\lambda} = 1,
\end{equation}
which means that the solution is asymptotically locally AdS$_5$ in global coordinates, with $L_5 = 2/g$.
Like we did for BPS attractors, we first solve the Killing spinor equation on the internal space with the ansatz for the gauge field,
but different projections are needed to get a full black string solution
\footnote{The corresponding projections for full black string solutions in the $\nn=8$ language are:
\begin{align}
\gamma^{23} \epsilon = \Gamma^{23} \epsilon = \Gamma^{14} \epsilon = \Gamma^{56} \epsilon, \\
\gamma^R \Omega \epsilon = \epsilon.
\end{align}}
: 
\begin{equation}
 \epsilon = \Gamma^{1} \epsilon= \gamma^R \gamma^{23} \epsilon = \gamma^R \Gamma^{45} \epsilon.
\end{equation}
These are three projections, so the full solutions preserve only two supercharges, while there is enhancement to four supercharges at the horizon.
Once we impose these projections the BPS equations \ref{BPSeq} 
reduce to a coupled system of three first order differential equations in the radial coordinate:
 \begin{align} \label{BPSfull}
  f'  = \frac{g}{3}(\sigma e^{-\lambda} + \frac{1}{2} e^{2\lambda}) & - \frac{1}{3 h^2} (p e^{\lambda}+ \tilde{p}e^{-2\lambda}), \\
  f \frac{h'}{h}  = \frac{g}{3}(\sigma e^{-\lambda} + \frac{1}{2} e^{2\lambda}) & + \frac{2}{3 h^2} (p e^{\lambda}+ \tilde{p}e^{-2\lambda}), \\
  f \lambda'  = \frac{g}{3}(\sigma e^{-\lambda} - e^{2\lambda}) & - \frac{1}{3 h^2} (p e^{\lambda}-2 \tilde{p}e^{-2\lambda}),
 \end{align}
plus an equation that determines the Killing Spinor to be:
 \begin{equation}
 \epsilon = \sqrt{f}\ \epsilon_0,
 \end{equation}
where $\epsilon_0$ is a constant spinor. 
It is now possible to interpolate between the two desired backgrounds with a numerical solution to the equations \ref{BPSfull},
and the result is a one dimensional family of BPS black strings. 
These solutions can be embedded in the numerical solutions constructed in \cite{Benini:2013cda} for the $\nn=2$ theory,
whose potential reduce to the $4^+$ potential after identifying the two scalars $\lambda_1 = \lambda_2 = \lambda$
\footnote{In the $\nn=8$ language the $\nn=2$ theory is derived as an $SO(2)\times SO(2)\times SO(2)$ invariant truncation,
whose scalar sector is described by two degrees of freedom collactable in the diaginal matrix:
\[ \Lambda_{\nn=2} = \left( \begin{array}{cccccc}
 \lambda_1 & & & & & \\
 & \lambda_1 & & & & \\
 & & \lambda_2 & & & \\ 
 & & & \lambda_2 & & \\
 & & & & -\lambda_1 - \lambda_2 & \\
 & & & & & -\lambda_1 - \lambda_2
 \end{array} \right),\]
which for $\lambda_1 = \lambda_2 = \lambda$ is the same matrix we used to describe the scalar sector $SO(4)\times SO(2)$ invariant truncation in section \ref{subsec:scalar}.   
}.

Reference \cite{Benini:2013cda} also provides a clear description of the dual two-dimensional field theory in the IR, i.e.\ on the horizon of the black strings. Put in the context of our present findings, this opens up an interesting possibility. In the previous section we saw that the BPS attractor geometry AdS$_3 \times$S$^2$ exists both in $4^+$ and in $4^0$ with the same amount and type of supersymmetry. It would therefore not be surprising if the dual CFT$_2$ in \cite{Benini:2013cda} is also relevant for black strings in Minkowski.

\section{Full analytic solutions}
\label{sec:analytic}
In section \ref{sec:exact} we presented a list of possible attractor geometries in maximal gauged supergravity,
and in section \ref{sec:RGflows} we focused on the subclass of BPS attractors, which can be connected numerically to infinity 
with a full supersymmetric solution. In this section we will consider a second important subset of near horizon geometries,
namely those that can be connected to infinity with a full analytic solution with constant scalars. 

It is very easy to derive the condition that determines this subset in the case of AdS$_5$ and dS$_5$ asymptotics,
where the vev of the scalar is fixed by its value at infinity:
\begin{equation}
 e^{\lambda} = 1.
\end{equation}
Once we impose this condition the scalar equations of motion are satisfied only if two gauge fields are turned on,
with the two charges related by:
\begin{align}
p^2  = \tilde{p}^2\ \ &\mbox{for} \ \ 3^{\sigma} , \\
p^2  = 4\tilde{p}^2\ \ &\mbox{for} \ \ 4^{\sigma},
\end{align}
in the case of magnetic solutions, or the analogue relations for electric solutions.


If we instead look for asymtotically flat solutions with constant scalar in the $4^0$ theory, 
the scalar equation of motion gives one single constraint:
\begin{equation}
 \tilde{p}^2= \frac{1}{4}e^{6\lambda} p^2 ,
\end{equation}
in the case of magnetic solutions, or the analogue relation for the electric ones.

We now give a list of all possible analytic solutions with constant scalar, organized accordingly to their asymptotics.

\subsection{in AdS$_5$}
\label{subsec:AdS5}
Full analytic solutions of this type can exist in the $3^+$ and $4^+$ truncations, 
which both have AdS$_5$ as maximally symmetric background.
The following black hole solutions also have the interpretation of holographic RG flows. 
\begin{itemize}
 \item \textbf{ Nonabelian Black Holes}
\\ Full analytic nonabelian solutions can be found only in the $3^+$ theory, where two nonabelian gauge fields can be turned on.
We start from  the following ansatz:
 \begin{align}
 & ds^2 = V(r)dt^2-V(r)^{-1}dr^2 - r^2 ds^2_{3,k},
\end{align}
and plug it into the equations of motion \eqref{eom}. We get a solution for:
\begin{equation}
 V(r)= k +\frac{r^2}{L_5^2}-\frac{M + 4 p^2 Log(r)}{r^2},
\end{equation}
where the mass parameter $M$ is free, while the length of AdS$_5$ is fixed:
\begin{equation}
 gL_5=2.
\end{equation}
For a generic value of $M$ we have distinct horizons, 
while in the extremal case $M$ is fixed and we get an AdS$_2 \times \Sigma_3^k$ geometry with background values for the parameters given by \eqref{NA33sol}.
The solution with spherical horizon $k=1$ was already presented in \cite{Okuyama:2002mh,Cvetic:2009id}, 
while for $k=-1$ we get a new solution corresponding to a $nonabelian\ topological\ black\ hole$.
In the case of flat horizon $k=0$ we also get a family of solutions which are Schwarzschild like black branes, 
since the field strength vanishes accordingly to the quantization of the Yang Mills charge \eqref{NAQ}.

 \item \textbf{Black Strings}
\\ Full analytic black string solutions are very rare. In fact, even though we were able to produce
a large variety of black string attractors which are listed in section \ref{subsec:solutions}, 
there exist only a single special case in which the near horizon geometry can be connected analytically to infinity.
We start from  the following ansatz:
 \begin{align}
 & ds^2 = V(r)dt^2-U(r)^{-1}dr^2 -V(r)dz^2 - r^2 ds^2_{2,k},
\end{align}
and plug it into the equations of motion \eqref{eom}. In both the $4^+$ and $3^+$ truncations we get a solution for:
\begin{align}
 V(r) & = \frac{L_5}{r} \left( 3k+ \frac{r^2}{L_5^2} \right)^{3/2}, \nonumber \\
 U(r) & = \left( \frac{r}{L_5} +k \frac{L_5}{3r}\right)^2,
\end{align}
where the AdS$_5$ length is fixed to:
\begin{equation}
 gL_5=2.
\end{equation}
The abelian charge gets quantized differently in the two truncations: 
\begin{align}
g^2p^2 & = \frac{k^2}{6}\ \ \mbox{for} \ \ 3^+ , \nonumber \\
g^2p^2 & = \frac{4 k^2}{9}\ \ \mbox{for} \ \ 4^+.
\end{align}
Notice that for $k=0$ we have vanishing magnetic charge and the metric becomes locally AdS$_5$. 
For $k=1$ we get a naked singularity, which is BPS in the $4^+$ case if both $p$ and $\tilde{p}$ are negative,
and correspond to the solution in the $\nn=2$ theory given in \cite{Chamseddine:1999xk}. 
Finally for $k=-1$ we get an extremal black string with a double horizon at $r=L_5/\sqrt{3}$. 
This solution is also BPS in the $4^+$ theory if both the magnetic charges are negative, 
and corresponds to the topological black string found in  \cite{Klemm:2000nj} for the $\nn=2$ theory. 
The near horizon geometry of this solution is AdS$_3 \times \mathbb{H}^2$, 
which correspond to a single point in the two dimensional parameter space of attractor geometries given in \eqref{42Asol} and \eqref{33Asol}. 

 \item \textbf{Black Holes}
\\ Full analytic solutions of this type can be found in both the $4^+$ and $3^+$ truncations, with two abelian gauge fields turned on.
Of course these solutions cannot be supersymmetric, as they correspond to the usual electrically charged Reissner-Nordstr\"{om} black holes 
in AdS \footnote{The supersymmetric limit of RN-AdS spacetimes is always a naked singularity, and the extremality bound is above the BPS one, 
therefore none of the regular RN-AdS black holes is supersymmetric.}, 
rewritten in the language of maximal gauged supergravity.
We start with  the following ansatz:
 \begin{align}
 & ds^2 = V(r)dt^2-V(r)^{-1}dr^2 - r^2 ds^2_{3,k},
\end{align}
and plug it into th equations of motion \eqref{eom}. In the $4^+$ case we get a solution for:
\begin{equation}
 V(r)= k +\frac{r^2}{L_5^2}-\frac{M }{r^2} + \frac{3}{4}\frac{Q^2}{r^4},
\end{equation}
while $3^+$ we get a slightly different expression:
\begin{equation}
 V(r)= k +\frac{r^2}{L_5^2}-\frac{M }{r^2} + \frac{2}{3}\frac{Q^2}{r^4},
\end{equation}
where the parameters $M$ and $Q$ are free, while the length of AdS$_5$ is determined to be:
\begin{equation}
 gL_5=2.
\end{equation}
In the extremal case we can fix both $M$ and $Q$ in terms of the horizon radius $R$. 
We get an AdS$_3 \times \Sigma_k^2$ geometry, 
whose background parameters are described by the solutions \eqref{42Esol} \eqref{33Esol} for $c=1$, $K=-1$, $\sigma = 1$.

 \end{itemize}

 \subsection{in Mink$_5$}
 \label{subsec:Mink5}

 Asymptotically flat solutions can only exist in the special $4^0$ theory which has a vanishing scalar potential.
 Being the bosonic lagrangian of $4^0$ the same as the ungauged supergravity, the two theories have the same spectrum of solutions,
 but different BPS sectors, as we showed in the previous section. Indeed we find a two parameter family of black strings that are BPS only at the horizon.
 
\begin{itemize}

 \item \textbf{Black Strings}
\\ We start from  the following ansatz:
 \begin{align} \label{40BS}
 & ds^2 = V(r)dt^2-V(r)^{-2}dr^2 -V(r)dz^2 - r^2 ds^2_{S^2},
\end{align}
where the function $V$ has the form:
\begin{equation}
 V(r)= 1 -\frac{R }{r}.
\end{equation}
We get a two parameter family of solutions. 
For ease of comparison with \eqref{42Asol} we keep the internal radius $R$ and the scalar $c$ are free parameters, 
while the magnetic charge is determined to be:
\begin{equation}
p^2 = \frac{ R^2}{c^2}.
\end{equation}
Observe that the metric \eqref{40BS} already describes the extremal case with two coincident horizons at $r=R$.
In the near horizon limit we get a 2 parameters family of backgrounds with geometry AdS$_3 \times S^2$,
which is precisely the full 2 dimensional moduli space we got in \eqref{42Asol} for $\sigma=0$, $K=-1$, $k=1$. 

Half of the near-horizon regions correspond to fully BPS solutions in the ungauged theory and the other half are half-BPS in the gauged theory. 
The full solutions are always non-BPS in gauged $4^0$ theory, 
but correspond to BPS and almost-BPS solutions from the point of view of ungauged supergravity, 
as already discussed in the previous section.

 \item \textbf{Black Holes}
\\ Of course in the $4^0$ theory we can also have static asymptotically flat black holes charged under an electric field, i.e.' the usual RN black holes in 5d, and those are non supersymetric in the gauged theory\footnote{As well known, RN black holes are BPS in the extremal limit in ungauged supergravity when $g=0$.}. We start with  the following ansatz:
 \begin{align}
 & ds^2 = V(r)dt^2-V(r)^{-1}dr^2 - r^2 ds^2_{S^3}, \end{align}
we get a solution for
\begin{equation}
 V(r)= 1 -\frac{M }{r^2}+\frac{Q^2}{4 r^4}.
\end{equation}
where $M$ and $Q$ are free. In the extremal case we can fix $M$ and  $Q$ in terms of the horizon radius $R$, 
to get a one parameter family of exact backgrounds AdS$_2 \times S^3$,
which can be obtained from \eqref{42Esol} for $K=-1$, $k=1$, $\sigma=0$.

 \end{itemize}

\subsection{in dS$_5$}
\label{subsec:dS5}

Solutions with dS$_5$ asymptotics can only exist in the $3^-$ theory and none of these can be supersymmetric.
These solutions are nevertheless interesting, as they are amongst the very few dS solutions that exist in supergravity. 
Also, in this section we describe a very peculiar phenomenon for near horizon geometries of black holes in de Sitter space:
they can be collected into a one parameter family of attractors M$_2 \times S^3$, where M$_2$ can be AdS$_2$, Mink$_2$ or dS$_2$ as the parameter changes. 

\begin{itemize}
 \item \textbf{ Nonabelian Black Holes}
\\ Full analytic solutions can be found when both the nonabelian gauge fields are turned on.
We start from  the following ansatz:
 \begin{align}
 & ds^2 = V(r)dt^2-V(r)^{-1}dr^2 - r^2 ds^2_{S^3}, 
\end{align}
and plug it into th equations of motion \eqref{eom}. We get a solution for:
\begin{equation}
 V(r)= 1 -\frac{r^2}{L_5^2}-\frac{M + 4 p^2 Log(r)}{r^2}.
\end{equation}
The length of dS$_5$ is fixed:
\begin{equation}
 gL_5=2\sqrt{2},
\end{equation} 
so the only free parameter is the mass $M$. The function $V(r)$ has a single horizon, the cosmlogical one,
so what we found is a one parameter family of $nonabelian$ $naked$ $singularities$ in de Sitter space.

 \item \textbf{Black Holes}
\\ A proper ansatz for black holes in e Sitter space is given by:
 \begin{align}
 & ds^2 = V(r)dt^2-V(r)^{-1}dr^2 - r^2 ds^2_{S^3}.
\end{align}
We can solve the equations of motion for:
\begin{equation}
 V(r)= 1 -\frac{r^2}{L_5^2}-\frac{M }{r^2} + \frac{2}{3}\frac{Q^2}{r^4},
\end{equation}
where the parameters $M$ and $Q$ are free, while the length of dS$_5$ is fixed:
\begin{equation}
 gL_5=2\sqrt{2}.
\end{equation}
Something very interesting happens in the extremal limit, where we get a one parameter family of M$_2\times S^3$ near horizon geometries,
where M$_2$ can be AdS$_2$, dS$_2$ or Mink$_2$ accordingly to the value of the parameters.
We first impose $V[R] = 0 = V'[R]$ and express $M$ and $Q$ in terms of the horizon radius $R$, 
then we get the following form for the extremal potential:
\begin{equation}
V_{ext}[r] = \frac{(r^2 - R^2)^2 ( 1 -\frac{g^2}{8} (r^2+2R^2) )}{r^4},
\end{equation}
Notice that this function always has a double horizon at $r=R$, 
plus a cosmologucal horizon at:
\begin{equation}
 r_c = \sqrt{ 8/g^2 - 2R^2}.
\end{equation}
As long as $ R^2 < \frac{8}{ 3g^2 }$ the cosmological horizon is outside the double horizon,
and the resulting near horizon geometry is AdS$_2 \times S^3$. 
Viceversa if $ \frac{8}{ 3g^2 } < R^2 < \frac{4}{g^2} $ the role of the two horizons gets inverted, 
and we get a dS$_2 \times S^3$ near horizon geometry.
Finally there is a special point in the parameter space corresponding to $R^2=\frac{8}{ 3g^2 }$ where we the cosmological horizon 
and the double horizon coincide, namely we get a $triple\ horizon$.
The resulting near horizon geometry is Mink$_2 \times S^3$,
as for a triple horizon also the second derivative of the potential vanishes $V''[R]=0$
and hence the Ricci tensor $R_{\mu \nu}$ vanishes at $r=R$.
 \end{itemize} 

\section{Summary}
\label{sec:summary}
In this paper we scanned for static solutions in a set of maximal gauged supergravities in five dimensions with compact and non compact gaugings.
For this purpose we introduced varius truncations of the full spectrum, 
whose scalar potentials admit either AdS$_5$, Mink$_5$ or dS$_5$ as maximally symmetric vacuum. 
\\ \indent We listed all possible attractor geometries for abelian and nonabelian black holes, strings, and rings.
These backgrounds often come in two parameter families
and in there should exist a black hole like solution connecting each point in the moduli space to an appropriate asymptotic at infinity.
\\ \indent We analyzed the special points in the moduli space that can be connected to infinity with
analytic solutions with constant scalars, and we summarize them in table $7$.
\begin{table}
\begin{center}\caption{Full analytic solutions}
\begin{tabular}{ |c|c|c| }
\hline
Black Object & Asymptotics & References \\ \hline
\multirow{3}{*}{BH} & AdS$_5$ & \ref{subsec:AdS5} \\
 & Mink$_5$ & \ref{subsec:Mink5} \\
 & dS$_5$ & \ref{subsec:dS5} \\ \hline
\multirow{3}{*}{BS} & AdS$_5$ & \ref{subsec:AdS5}, \cite{Klemm:2000nj} \\
 & Mink$_5$ & \ref{subsec:Mink5}, \cite{Compere:2010fm} \\
 & dS$_5$ & ? \\ \hline
\multirow{2}{*}{NABH} & AdS$_5$ & \ref{subsec:AdS5}, \cite{Cvetic:2009id} \\
 & dS$_5$ & \ref{subsec:dS5} \\
\hline
\end{tabular}
\end{center}
\end{table}
Apart from the full analytic solution for black strings in de Sitter space,
all the other possibilies have been covered.
\\ \indent A special sector in the moduli space is given by the BPS near horizon geometries, which were also listed exhaustively.
It is possible to find a numerical solution to the BPS equations connecting these types of backgrounds to either AdS$_5$ or Mink$_5$ in the $4^+$ and $4^0$
theories respectively. Interpolating BPS solutions of this type have been constructed in \cite{Nieder:2000kc,Naka:2002jz} for non abelian black holes,
and in \cite{Benini:2013cda} for black strings in the less supersymmetric $\mathcal{N}=2$ truncation. 
\\ \indent We expect that all other non-BPS near horizon geometries can be connected numerically to an appropriate asymptotic at infinity through the equations of motion \eqref{eom},
for an ansatz with running scalars.  

\section*{Acknowledgments}
We would like to thank B.\ de Wit, N.\ Halmagyi, S.\ Katmadas and C.\ Klare for interesting discussions, and to A.\ Tomasiello and A.\ Zaffaroni for initial motivation and collaboration and numerous discussions. We are supported in part by INFN, by the MIUR-FIRB grant RBFR10QS5J ``String Theory and Fundamental Interactions'', and by the MIUR-PRIN contract 2009-KHZKRX.






\providecommand{\href}[2]{#2}
\end{document}